\PassOptionsToPackage{pdfpagelabels=false}{hyperref} 

\documentclass[fleqn,usenatbib]{mnras}

\usepackage{newtxtext,newtxmath}

\usepackage[T1]{fontenc}
\usepackage{ae,aecompl}
\usepackage[dvipsnames]{xcolor}
\usepackage{color}
\usepackage{multirow}

\usepackage{graphicx}	
\usepackage{amsmath}	
\usepackage{amssymb}	
\usepackage{siunitx}
\newcommand{\um}[1]{\SI{#1}{\micro\meter}}

\title[Comparison between EAGLE and DustPedia]{Reproducing the Universe: a comparison between the EAGLE simulations and the nearby DustPedia galaxy sample}

\author[Ana Tr\v{c}ka et al.]{Ana Tr\v{c}ka$^{1}$\thanks{E-mail: ana.trcka@ugent.be},
Maarten Baes$^{1}$,
Peter Camps$^{1}$,
Sharon E. Meidt$^{1}$,
James Trayford$^{2}$, \newauthor
Simone Bianchi$^{3}$,
Viviana Casasola$^{4,3}$,
Letizia P. Cassar\`{a}$^{5}$,
Ilse De Looze$^{1,6}$,\newauthor
Pieter De Vis$^{7}$,
Wouter Dobbels$^{2}$,
Jacopo Fritz$^{8}$,
Maud Galametz$^{9}$,
Fr\'{e}d\'{e}ric Galliano$^{9}$,\newauthor
Antonios Katsianis$^{10,11}$,
Suzanne C. Madden$^{9}$,
Aleksandr V. Mosenkov$^{12,13}$,\newauthor
Angelos Nersesian$^{1,14,15}$,
S\'{e}bastien Viaene$^{1,16}$,
 and Emmanuel M. Xilouris$^{14}$
\\
$^{1}$Sterrenkundig Observatorium, Universiteit Gent, Krijgslaan 281, B-9000 Gent, Belgium\\
$^{2}$ Leiden Observatory, Leiden University, PO Box 9513, NL-230 0 RA Leiden, The Netherlands\\
$^{3}$ INAF - Osservatorio Astrofisico di Arcetri, Largo E. Fermi 5, I-50125, Florence, Italy\\
$^{4}$ INAF - Istituto di Radioastronomia, Via P. Gobetti 101, I-40129, Bologna, Italy\\
$^{5}$ INAF – Istituto di Astrofisica Spaziale e Fisica Cosmica, Via Alfonso Corti 12, 20133 Milan, Italy\\
$^{6}$ Department of Physics and Astronomy, University College London, Gower Street, London WC1E 6BT, UK\\
$^{7}$ School of Physics and Astronomy, Cardiff University, The Parade, Cardiff CF24 3AA, UK\\
$^{8}$ Instituto de Radioastronom\'{i}a y Astrof\'{i}sica, UNAM, Campus Morelia, A.P. 3-72, C.P. 58089, Mexico\\
$^{9}$ AIM, CEA, CNRS, Universit\'{e} Paris-Saclay, Universit\'{e} Paris Diderot, Sorbonne Paris Cit\'{e}, F-91191 Gif-sur-Yvette, France\\
$^{10}$ Tsung-Dao Lee Institute, Shanghai Jiao Tong University, Shanghai 200240, China \\
$^{11}$ Department of Astronomy, Shanghai Key Laboratory for Particle Physics and Cosmology, Shanghai Jiao Tong University, Shanghai200240, China\\
$^{12}$ Central Astronomical Observatory of RAS, Pulkovskoye Chaussee 65/1, 196140, St. Petersburg, Russia\\
$^{13}$ St. Petersburg State University, Universitetskij Pr. 28, 198504, St. Petersburg, Stary Peterhof, Russia\\
$^{14}$ National Observatory of Athens, Institute for Astronomy, Astrophysics, Space Applications and Remote Sensing, Ioannou Metaxa
and Vasileos Pavlou\\ GR-15236, Athens, Greece\\
$^{15}$ Department of Astrophysics, Astronomy \& Mechanics, Faculty of Physics, University of Athens, Panepistimiopolis, GR-15784
Zografos, Athens, Greece\\
$^{16}$ Centre for Astrophysics Research, University of Hertfordshire, College Lane, Hatfield, AL10 9AB, UK
}
\date{Accepted XXX. Received YYY; in original form ZZZ}

\pubyear{2019}

\begin{document}
\label{firstpage}
\pagerange{\pageref{firstpage}--\pageref{lastpage}}
\maketitle

\begin{abstract}
We compare the spectral energy distributions (SEDs) and inferred physical properties for simulated and observed galaxies at low redshift. We exploit UV-submillimetre mock fluxes of $\sim 7000$ z=0 galaxies from the EAGLE suite of cosmological simulations, derived using the radiative transfer code \textsc{skirt}. We compare these to $\sim 800$ observed galaxies in the UV-submillimetre range, from the DustPedia sample of nearby galaxies. To derive global properties, we apply the SED fitting code \textsc{cigale} consistently to both data sets, using the same set of $\sim 80$ million models. The results of this comparison reveal overall agreement between the simulations and observations, both in the SEDs and in the derived physical properties, with a number of discrepancies. The optical and far-infrared regimes, and the scaling relations based upon the global emission, diffuse dust and stellar mass, show high levels of agreement. However, the mid-infrared fluxes of the EAGLE galaxies are overestimated while the far-UV domain is not attenuated enough, compared to the observations. We attribute these discrepancies to a combination of galaxy population differences between the samples, and limitations in the subgrid treatment of star-forming regions in the EAGLE-\textsc{skirt} post-processing recipe. Our findings show the importance of detailed radiative transfer calculations and consistent comparison, and provide suggestions for improved numerical models.
\end{abstract}




\section{Introduction}

Despite the fact that over the last decades our knowledge of galaxy formation and evolution has improved substantially, we still have only a fragmentary understanding of all the complex and coupled physical phenomena that shape galaxies. Numerical simulations of galaxy formation and evolution \citep[][and references therein]{Vogelsberger2019} are a needed and valuable tool to alleviate these difficulties, provided that they are able to reproduce galaxy populations that, in various aspects, resemble the ones found in the real Universe. 
Therefore, it is necessary to compare the simulated and observed objects in order to test the models and also to fine-tune the subgrid parameters.

In recent years, the power of cosmological hydrodynamical simulations increased  immensely \citep[e.g.][]{Vogelsberger2014, Schaye2015, Pillepich2018, Dave2019}. They are able to reproduce galaxy properties and scaling relations that were not used for calibration, including hydrogen content, colours, morphology and properties of satellite galaxies \citep{Lagos2015, Sales2015, Trayford2015, Bahe2017, Crain2017, Nelson2018, Diemer2019}. However, comparing simulations to observations is not trivial since the output from simulations (e.g. stellar mass, star formation rates (SFRs), metallicity, etc.) usually is not directly comparable to observational data (e.g. fluxes at various broadbands). 

Commonly, a comparison is made in the physical realm, which involves adopting different assumptions, tracers and recipes to calculate physical properties from the observed light. This approach can introduce systematics and uncertainties, even when deriving relatively simple properties such as stellar masses and SFRs \citep{RosaGonzalez2002,Mitchell2013,Guidi2015}, and even when the same method of derivation is used (e.g. spectral energy distribution (SED) fitting) but different codes \citep{Pappalardo2016, Hunt2019}. 

An alternative approach is to compare directly in the observed flux space. In contrast with the previous method, this one requires intensive treatment of the simulation data in order to obtain realistic mock observations. This method is needed if one wishes to investigate galaxy morphology \citep{Dickinson2018, Rodriguez-Gomez2019, Bignone2019}, or extract galaxy colours \citep{Trayford2015, Trayford2017, Nelson2018}, or the whole SED \citep{Camps2017, Liang2019, Ma2019, Katsianis2020}.

To obtain the most realistic mock observations of simulated galaxies, aside from stars and gas, it is necessary to also include interstellar dust in the modelling. During the last decades, we have grown to understand the importance of cosmic dust as a powerful medium for distorting stellar light in galaxies \citep{Calzetti1994, Galliano2018}. Dust reprocesses more than $30\%$ of stellar radiation of typical star-forming galaxies, entirely reshaping galaxy SEDs through processes of absorption, scattering and then re-emission at longer wavelengths \citep{Popescu2002, Skibba2011, Viaene2016, Bianchi2018}. Thus, despite its low mass fraction of less than $1\%$ of the interstellar medium mass \citep{RemyRuyer2014}, dust is a crucial ingredient in the Universe. Nevertheless, dust is still rarely modelled in cosmological simulations.

One approach adopted is to incorporate dust creation, growth, destruction and dynamics into simulations directly \citep{McKinnon2016, McKinnon2017, Aoyama2018, Aoyama2019, Hou2019, Dave2019}. This, however is a very computationally expensive method, involving many processes that remain poorly understood. A simpler method, that can be applied more easily to large-scale cosmological simulations, is to model dust based on information on gas and stars from the simulation \citep{Camps2016, Trayford2017, Liang2018, Narayanan2018, Cochrane2019, Liang2018, Ma2019, Rodriguez-Gomez2019}, and then perform radiative transfer in post processing.

EAGLE \citep{Schaye2015, Crain2015} is a suite of state-of-the-art cosmological hydrodynamical simulations. Since these simulations do not include dust in the modelling, assumptions are needed to run the radiative transfer simulations. \citet{Camps2016} and \citet{Trayford2017} introduced a radiative transfer post-processing procedure of the EAGLE simulations using the \textsc{skirt} code \citep{Baes2011}, and they tested how well the coupling of EAGLE and \textsc{skirt} recreates infrared (IR) and submillimetre as well as ultraviolet (UV) and optical observations of the Local Universe, respectively. 
They calibrated the parameters associated with the post-processing procedure using three different scaling relations to achieve the best agreement between a sub-sample of around 300 galaxies from the Herschel Reference Survey \citep[HRS:][]{Boselli2010, Cortese2012} and a K-band luminosity matched sample of EAGLE galaxies. Since these studies showed broad agreement with the observations, \citet{Camps2017} enriched the public EAGLE database \citep{McAlpine2016} with the mock fluxes for most of the EAGLE galaxies, based on the same modelling prescriptions as before.

There are, however, a couple of caveats. The HRS sample is limited in size, and centred on the Virgo Cluster and thus may be less representative for the general galaxy population. Also, dust and stellar masses, which \citet{Camps2016} used for the calibration, were derived adopting simple recipes using only SPIRE data and the SDSS \textit{g} and \textit{i} bands, respectively, so in total only 5 bands. Therefore, the calibration did not take into account data at far UV (FUV) and mid infrared wavelengths (MIR), which are crucial for the calculation of physical parameters such as stellar mass and SFR and which depend critically on the properties and distribution of the dust \citep{Bell2003, Kennicutt2012}.

In their recent work, \citet{Baes2019} argue that the described EAGLE-\textsc{skirt} coupling broadly reproduces the Galaxy And Mass Assembly \citep[GAMA:][]{Driver2011,Liske2015} cosmic SED\footnote{Total energy in a cosmologically representative volume at different wavelengths.} at $z=0$ \citep{Andrews2017}. However, the comparison shows tension at UV wavelengths, revealing that the attenuation by the EAGLE-\textsc{skirt} model at these wavelengths is underestimated. This indicates discrepancy in the galaxy populations between samples, or the need to improve the radiative transfer post-processing recipe. The analysis of the cosmic SEDs alone, however, is insufficient to fully understand the cause and the treatment of the potential issue. To achieve this one has to analyse individual galaxies.

The aim of this paper is to compare the EAGLE simulated galaxies to observed galaxies in the nearby Universe, and in particular to verify the calibration of the EAGLE-\textsc{skirt} procedure, developed by \cite{Camps2016} and \citet{Trayford2016}, and extended by \citet{Camps2017}. We perform the analysis at low redshifts, since nearby galaxies benefit from higher signal to noise data, enabling detailed characterisation across wavelength. As a comparison sample, we use DustPedia, the largest sample of nearby galaxies with matched aperture photometry in more than 40 bands from UV to millimetre wavelengths \citep{Clark2018}. The main advantages over the HRS sub-sample, which was used in the original EAGLE-\textsc{skirt} calibration, are the larger spread in environment and almost three times higher number of galaxies \citep{Davies2017}. In this paper, we exploit advantages of both comparative approaches, by comparing the samples in the two domains: of the observed fluxes and of physical properties. We use the fluxes from the post-processing of the EAGLE galaxies, which together with the observed DustPedia fluxes we treat in the same SED fitting environment of the \textsc{cigale} code \citep{Boquien2019}. Taking into account information across the entire UV-submillimetre wavelength range, and with the same assumptions and model parameters for both simulations and observations, we derive physical properties, and therefore compare these samples in a consistent way. 

We organise the paper as follows. In Sect. \ref{sec:methods}, we briefly review the EAGLE simulations, our \textsc{skirt} post-processing procedure, the DustPedia sample, and the \textsc{cigale} SED fitting procedure. In Sect. \ref{sec:results}, we perform the comparison between EAGLE-\textsc{skirt} simulations and the DustPedia sample of nearby galaxies. In Sect. \ref{sec:llplots}, we analyse relations between the observational and mock fluxes, prior to SED fitting. This provides some insight into differences between the samples which we analyse in more depth when comparing the SEDs in Sect. \ref{sec:SEDs}. 
Physical properties derived from our \textsc{cigale} fitting are presented in Sect. \ref{sec:proxies}. Our results are discussed in Sect. \ref{sec:diss} and summarised in Sect. \ref{sec:sum}.

\section{Methods}
\label{sec:methods}

\subsection{The EAGLE simulation suite}
\label{sec:EAGLE} 
We summarise the characteristics of the EAGLE simulations relevant to our study, but refer to \citet{Schaye2015} and \citet{Crain2015} for full details. The EAGLE suite consists of cosmological hydrodynamical simulations with different resolutions, subgrid models and a range of box sizes up to 100 comoving\footnote{The length does not change due to space expansion.} Mpc on a side. The simulations were performed using a modified version of the N-Body Tree-PM smoothed particle hydrodynamics code \textsc{gadget} 3 \citep{Springel2005}. The adopted cosmology is $\Lambda$CDM with parameters constrained by the Planck mission \citep{PlanckCollaboration2013}. The assumed stellar initial mass function (IMF) is that of \citet{Chabrier2003}.

All simulations incorporate subgrid models for radiative cooling, star formation, stellar evolution and enrichment, black hole seeding and growth, stellar and active galactic nuclei (AGN) feedback. Free subgrid parameters were calibrated to reproduce the observed $z = 0.1$ galaxy stellar mass function, galaxy size, and the relation between the black hole and stellar mass of galaxies.  

For this study, we are focusing on the redshift $z=0$ and on two EAGLE simulations: The reference Ref-L100N1504 (hereafter called the Ref-100 simulation) and Recal-L025N0752 (hereafter Recal-25). Recal-25 has higher resolution and, appropriately, a different set of subgrid recipes than Ref-100. Hence, regarding `weak' convergence (recalibration of subgrid physics as a consequence of a changed resolution, see Sect. 2.2 in \citealt{Schaye2015}), these simulations are comparable. The main properties of Ref-100 and Recal-25 are listed in Table~\ref{tab:eagle}. The last two columns represent mass and spatial resolutions. 

\begin{table}
	\centering
	\caption{List of relevant properties: name; box size of the simulation in comoving Mpc (L); number of particles (N); gas particle initial mass ($m_g$); maximum proper gravitational softening length ($\epsilon_{\mathrm{prop}}$).}
	\label{tab:eagle}
	\begin{tabular}{lcccc} 
		\hline
		Simulation name (label) & L & N & $m_g$ & $\epsilon_{\mathrm{prop}}$\\
		& cMpc & & $\mathrm{M_\odot}$ & kpc\\\hline
		Ref-L1001504 (Ref100) & 100 & $1504^{3}$ & $1.81\times 10^6$ & 0.70\\
		Recal-L025N0752 (Recal25) & 25 & $752^{3}$ & $2.26\times 10^5$ & 0.35\\
		\hline
	\end{tabular}
\end{table}

\subsection{Post-processing EAGLE with \textsc{skirt}}
\label{sec:SKIRT} 

\textsc{skirt} is a state-of-the-art Monte Carlo radiative transfer code \citep{Baes2003, Baes2011, Camps2015} that incorporates all relevant processes between dust and radiation in a galaxy (absorption, scattering, dust emission and dust self-absorption). One of its features is the capability to calculate mock observations from the snapshot data of a hydrodynamical simulation. The method is described in \citet{Camps2016} and \cite{Trayford2017}. In order to expand these studies to higher redshifts, \citet{Camps2017}, employed a slightly modified method, calculating mock fluxes for all galaxies above a stellar mass threshold of $10^{8.5}~ \mathrm{M_{\odot}}$ for 6 EAGLE simulations described in \citet{Schaye2015}. Here, we list the most relevant aspects of the method and the galaxy sample. 

\citet{Camps2016} and \citet{Trayford2017} chose the value for the stellar mass threshold of $10^{8.5}~\mathrm{M_{\odot}}$ to have at least 100 star particles (below this value, sampling effects become dominant). They extracted gas and star particles enclosed within 30 kpc (to approximate a Petrosian aperture), as suggested in \citet{Schaye2015} and \citet{Crain2015}. Galaxies at redshift $z=0$ were placed at 20 Mpc. In this study, for the intermediate resolution run Ref-100 we impose a higher stellar mass threshold of $10^{9}~\mathrm{M_{\odot}}$, to ensure our sample contains sufficiently resolved galaxies. With the additional dust cut explained below, the minimum number of stellar particles for Ref-100 (Recal-25) is 1100 (2576).

For each stellar particle, the SED was acquired from the \textsc{galexev} library \citep{Bruzual2003}, based on age, metallicity and initial mass of the particle.

Since dust is not modelled in the EAGLE simulations, post-processing of the input data (stellar and gas particles) was required. The model includes two dust sources: star-forming regions (not resolved in the EAGLE simulations), and diffuse dust. The AGN effects were not modelled. 

To acquire `star-forming particles' from the EAGLE data, \citet{Camps2016} and \citet{Trayford2017} first select star particles younger than $100~\mathrm{Myr}$ and star-forming gas particles. 
They re-sample each star-forming particle and assign formation times based on the SFR of the parent particle: those with formation times lower than $10~ \mathrm{Myr}$ stay in the star-formation particle bin, those with higher ones are moved to a star particle bin, whereas those not formed are moved to a gas particle bin. 
Then an SED from the \textsc{mappings-iii} family \citep{Groves2008} was assigned to each star-forming particle, based on its SFR, metallicity, pressure of the interstellar medium, compactness and $f_{\mathrm{PDR}}$, representing the covering fraction of the photo-dissociation regions (PDRs). 

The diffuse dust distribution was derived from the distribution of gas. The assumed dust model was from \citet{Zubko2004}, which consists of bare graphite and silicate grains, and polycyclic aromatic hydrocarbon (PAH) molecules, and uses solar interstellar medium (ISM) abundances. 
The dust mass was derived from cool or star-forming gas, and depends on the fraction of metals in dust $f_{\mathrm{dust}}$\footnote{Defined as $ \frac{\rho_{\mathrm{dust}}}{Z\rho_{\mathrm{gas}}},$ with $Z$, $\rho_{\mathrm{dust}}$ and $\rho_{\mathrm{gas}}$  as metallicity, dust and gas density, respectively.}.

The post-processing pipeline had two free parameters:$f_{\mathrm{dust}}$ and $f_{\mathrm{PDR}}$. \citet{Camps2016} selected the values for these parameters based on three scaling relations: the submillimetre colour diagram, specific dust mass ($M_{\mathrm{dust}}/M_{\mathrm{star}}$) versus stellar mass, and $M_{\mathrm{dust}}/M_{\mathrm{star}}$ versus the NUV-\textit{r} colour. 
The comparison was performed between galaxies from the HRS sub-sample and a matched sample of about 300 EAGLE galaxies. 
The adopted value of the covering fraction is $f_{\mathrm{PDR}}=0.1$ (below the reference value of \citealt{Jonsson2010}). 
They also adopt a metal fraction $f_{\mathrm{dust}}$ of 0.3 \citep{Dwek1998, Brinchmann2013}. 
Following, \citet{Camps2017} slightly changed the procedure. Firstly, they incorporated the process of dust self-absorption. Secondly, since the number of EAGLE galaxies is rather large, the radiative transfer procedure only included the calculation of the broadband fluxes, and \citet{Camps2017} did not generate resolved images for each individual galaxy. Considering the images are not produced,  the effects of the observational limits (e.g. surface brightness sensitivity limits of the telescope), are not accounted for. \citet{Camps2017} applied the procedure on 3 different angles: face-on, edge-on and random. In this study we use only the random angle which corresponds to the original galaxy orientation in the simulation. This way we mimic the random orientation in the observed galaxy sample.

The EAGLE mock data used in this paper (dust-attenuated and dust emission fluxes) are from \citet{Camps2017} and they are extracted from the public database\footnote{http://icc.dur.ac.uk/Eagle/database.php} \citep{McAlpine2016}. 

\begin{figure*}
	\includegraphics[width=2\columnwidth]{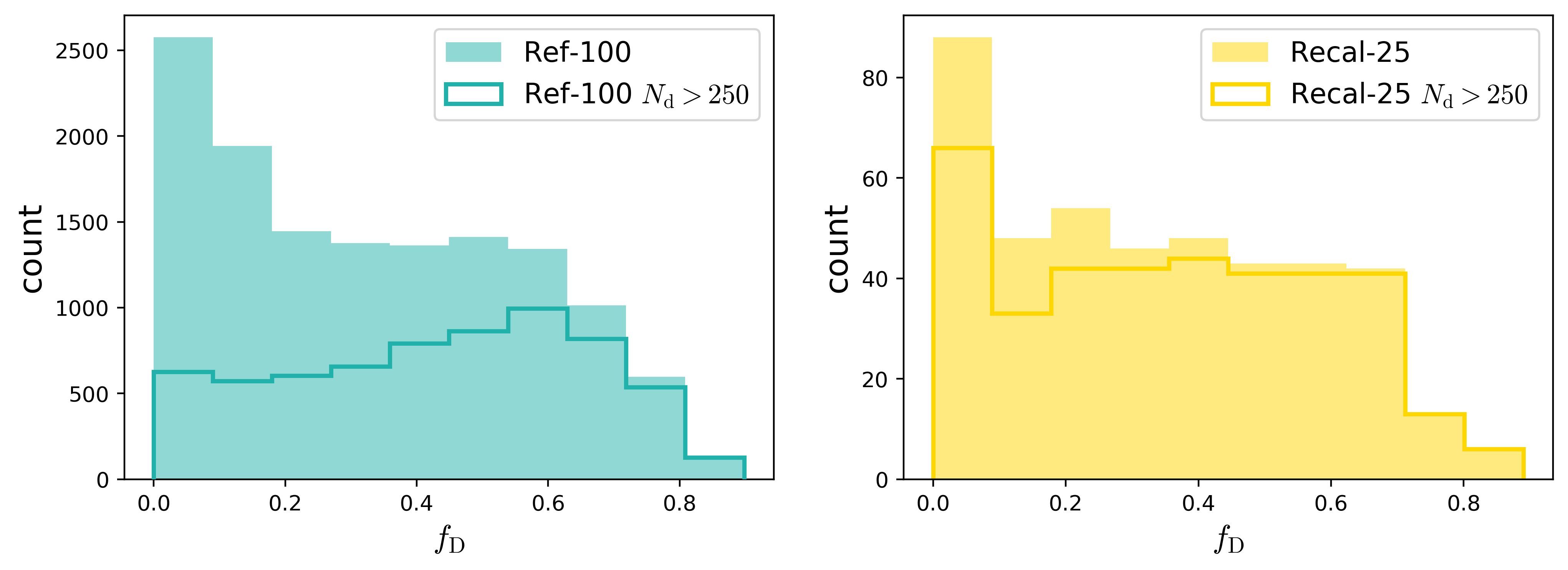}
    \caption{Distribution of the disc stellar mass fraction for EAGLE at $z=0$. Left panel shows results for Ref-100 ($M_{\mathrm{star}}>10^{9}~\mathrm{M_{\odot}}$) and right for Recal-25 ($M_{\mathrm{star}}>10^{8.5}~\mathrm{M_{\odot}}$). Filled histograms indicate the original EAGLE sample while outlined indicate our sample with resolved dust i.e. with the number of dust particles above 250.}
    \label{fig:morphhist}
    
\end{figure*}

It was already indicated in \citet{Camps2016}, and then confirmed in \citet{Camps2017} that the post-processing procedure produces unphysically low dust temperatures, for a fraction of simulated galaxies (see Fig. 3 in \citealt{Camps2017}). The cause of this is that these galaxies have insufficiently resolved dust distribution to characterize a realistic dust-to-stellar geometry. 
Therefore, from our samples we exclude galaxies that have less than 250 dust particles, as suggested by \citet{Camps2017}. As a consequence our Ref-100 sample is $50\%$ smaller and Recal-25 is $15\%$ smaller than the original EAGLE samples. To understand which galaxy type is mostly affected by this dust cut, we
exploited the morphology data of the EAGLE galaxies \citep{Trayford2019}. We analysed the disc stellar mass fraction $f_\mathrm{D}$, a parameter defined as $1~ - $ bulge-to-total mass ratio, with the bulge defined as twice the mass of the counter-rotating stellar particles \citep{Abadi2003}. 
In Fig. \ref{fig:morphhist}, we compared distributions of $f_\mathrm{D}$ of our sample and the whole EAGLE sample ($\log M_{\mathrm{star}}/\mathrm{M_\odot}>8.5 ~ (9)$ for Recal-25 (Ref-100), and at $z=0$ for both simulation runs).
Not surprisingly, our Ref-100 sample lacks most of the elliptical galaxies (those with $f_{D}\lesssim 0.5$), which are mainly red galaxies with low SFR (see Fig. 4 in \citealt{Camps2017}).

\subsection{DustPedia}
\label{sec:dustpedia} 
DustPedia \citep{Davies2017} is a European project initiated in order to improve our knowledge of cosmic dust and its role in the Local Universe. The DustPedia sample contains 875 nearby galaxies, observed with \textit{Herschel's} PACS or SPIRE instruments \citep{Pilbratt2010,Poglitsch2010,Griffin2010}. 
For a nearby, but still diverse, sample of galaxies populating different environments, objects are selected to have radial velocities below $3000\, \mathrm{km\; s^{-1}}$. 
Additionally, all galaxies have at least $5\sigma$ WISE \um{3.4} flux detection. In addition to the  \textit{Herschel} data, the DustPedia database\footnote{http://dustpedia.astro.noa.gr/} also includes data from GALEX \citep{Morrissey2007}, SDSS \citep{York2000}, 2MASS \citep{Skrutskie2006}, WISE \citep{Wright2010}, \textit{Spitzer} \citep{Werner2004},  \textit{Planck} \citep{Planck2011} and IRAS \citep{Neugebauer1984}.

\citet{Clark2018} presented an aperture-matched photometry of the whole DustPedia galaxy sample, except for \textit{Planck} and IRAS because of their poor resolution. The photometry for the EAGLE galaxies is derived in a similar manner with the same aperture for all bands. The average aperture for DustPedia is 17.6 kpc, which is below the value adopted for the EAGLE galaxies of 30 kpc. However, this difference is not affecting our analysis because twice the average stellar half mass radius for the Ref-100 (Recal-25) simulation is  11 kpc (8.5 kpc). This means most of the galaxies in our sample would be captured by a 17.6 kpc aperture. Additionally, the small DustPedia apertures correspond to low stellar mass galaxies. If we inspect only the galaxies with $M_{\mathrm{star}}>10^{8.5} \mathrm{M_\odot}$ (as those are more comparable with our EAGLE samples), we have a mean aperture of 18.8 kpc.

We do not include galaxies that have contamination from a nearby source, imagery artefacts or lack essential bands to constrain the SED fitting (e.g. bands in the optical and FIR). Our final DustPedia sample contains 814 galaxies.
Basic information about the 3 galaxy samples are shown in Table \ref{tab:DE}.

\begin{table}
	\centering

	\caption{Main characteristics of the two EAGLE galaxy samples and the DustPedia sample: the total number of galaxies, the aperture size, the distance from the galaxies, and the number of available bands for all the samples. For the EAGLE samples, the values are the same for every galaxy, while we provide the $16\% - 84\%$ range and mean values for the DustPedia galaxies.}
	\label{tab:DE}
	\begin{tabular}{lcccc} 
		\hline
		Galaxy sample & $N_\mathrm{gal}$ & Aperture  & Distance & $N_{\mathrm{bands}}^{(a)}$\\
				&  & kpc & Mpc & \\\hline
		
		Ref-100 & 6593 & $30$ & $20$ & 29\\\hline
		Recal-25 & 369 & $30$ & $20$ & 29\\\hline
		 \multirow{2}{*}{DustPedia} &  \multirow{2}{*}{814} & [7.7, 26.3] & [12, 33] & [16, 24]\\
		 &   & <17.6> & <21.5> & <20>\\\hline
		 	\multicolumn{5}{l}{$(a)$ Only bands with a positive flux are included.}
	\end{tabular}

\end{table}

\subsection{CIGALE}
\label{sec:cigale} 

We rely on the \textsc{cigale} fitting code (version 0.12.1) \citep{Noll2009, Boquien2019} to perform the SED fitting and derive physical properties such as the stellar and dust mass, SFR, dust luminosity etc. \textsc{cigale} incorporates stellar, nebular, AGN and dust emission and dust attenuation. It contains an implementation of a delayed and truncated star-formation history (SFH) \citep{Ciesla2016}, \citet{Bruzual2003} simple stellar population (SSP) libraries using \citet{Salpeter1955} IMF, modified \citet{Calzetti2000} attenuation law and the THEMIS \citep{Jones2017} dust model. The resulting library has of over 80 million model SEDs.
For details on the selection of modules, fitting process, parameter space and results for the DustPedia galaxies, we refer to \citet{Nersesian2019} and their Table 1. 

The presence of an AGN or a strong jet can affect the SED of a host galaxy, with the highest contribution in the IR part of the spectrum \citep[][Viaene et al., in preparation]{Mullaney2011, Xilouris2004}. The DustPedia sample contains 19 galaxies with a high probability of hosting an AGN \citep{Bianchi2018} and 4 jet-dominated galaxies \citep{Nersesian2019}. Since these galaxies account for only a small percentage of the sample, the additional \textsc{cigale} modules were not included to avoid computational costs \citep{Nersesian2019}. Since the AGNs are not modelled in the post-processing of EAGLE, we use also only non-AGN templates for the EAGLE galaxies. However, these DustPedia sources will be indicated separately on our plots.

The EAGLE database already contains stellar mass and SFR information but we choose to re-derive these properties using \textsc{cigale} to compare the simulated and real samples in a consistent approach. 

We note that a different IMF was used for the creation of the EAGLE simulations (as described in Sect. \ref{sec:EAGLE}), and for the SED fitting. The differences in the IMFs do not affect the comparisons we perform here, except the absolute values of the stellar mass and SFR, which are $\sim 0.2$ dex higher when derived with the use of the \citet{Salpeter1955}, than with the \citet{Chabrier2003} IMF.

An additional caveat is that the dust model adopted in the post-processing (\citet{Zubko2004}, see Sect. \ref{sec:SKIRT}), differs from the one used in the SED fitting (THEMIS). The main differences between the two models  are: (1) the FIR-submillimetre emissivity is about a factor of two higher for the THEMIS model; (2) a couple of aromatic bands around \um{20} are accounted for only in the model by \citet{Zubko2004}; (3) the strength of the aromatic features relative to the FIR pick emission is two times higher in the THEMIS model compared to the model by \citet{Zubko2004}, see  Fig. 4b of \citet{Galliano2018}. We discuss in later sections how these differences affect our comparison.   
We chose not to change the IMF, the dust model or any other parameter in the SED modelling, for the consistency, and easier comparison with the previous studies of the DustPedia galaxies \citep[][etc.]{Bianchi2018, Clark2018, DeVis2019, Nersesian2019,  Dobbels2020, Casasola2020}. Additionally, the same \textsc{cigale} fitting procedure for the EAGLE galaxies is already published in \citet{Baes2019}.  For the same reason we also use the same set of bands (i.e. all but \textit{Planck} bands, since they are not included in the public EAGLE database. We do not expect their absence would affect the SED fitting since every EAGLE galaxy has the FIR/submm region covered with all SPIRE bands.).

\begin{figure}
	\includegraphics[width=\columnwidth]{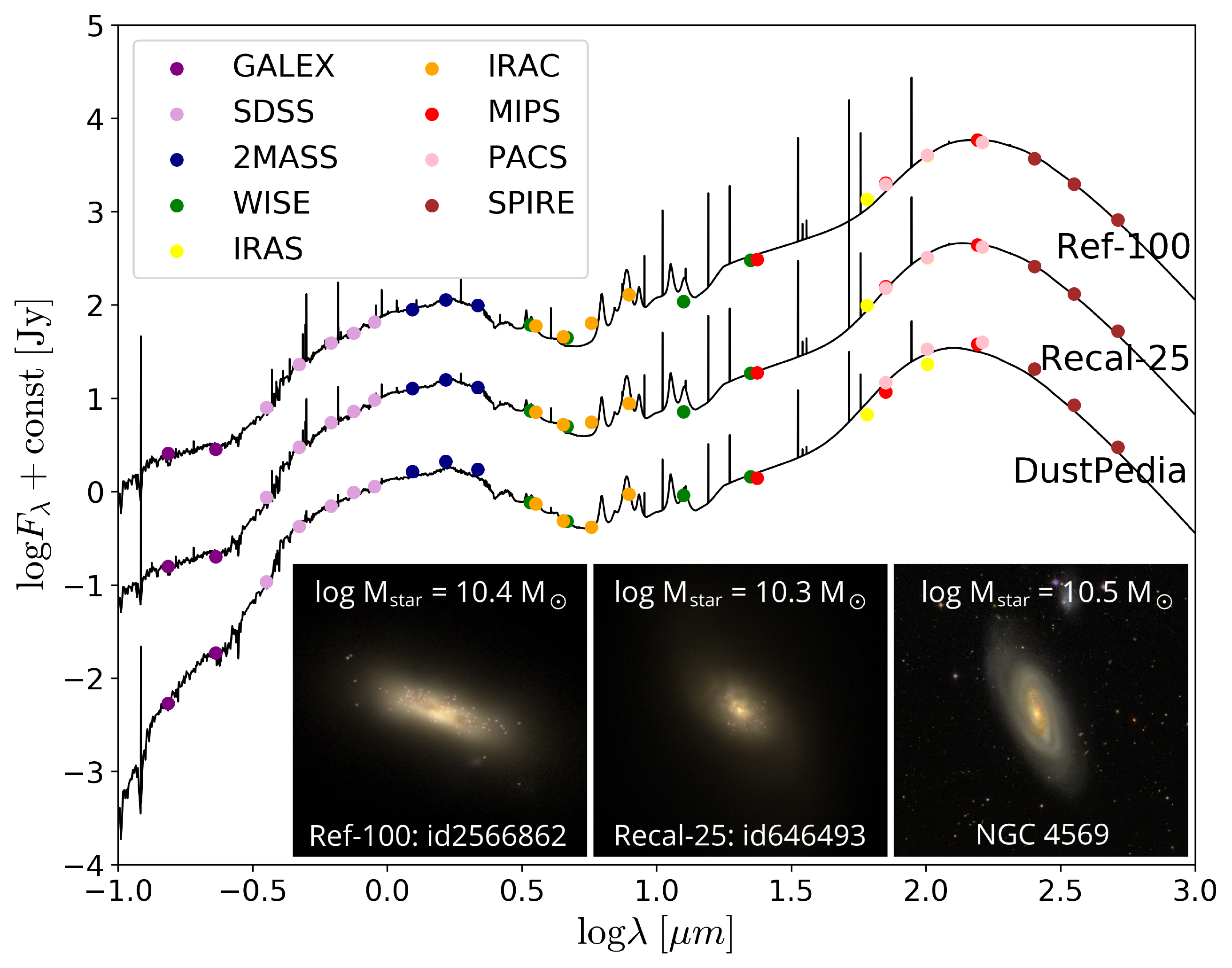}
    \caption{Example of SED data and fits derived by \textsc{cigale} of an arbitrarily selected galaxy for each galaxy sample. Images in \textit{gri} bands and the stellar masses of the same galaxies are also shown.}
    \label{fig:ECsed}
\end{figure}

\begin{figure}
	\includegraphics[width=\columnwidth]{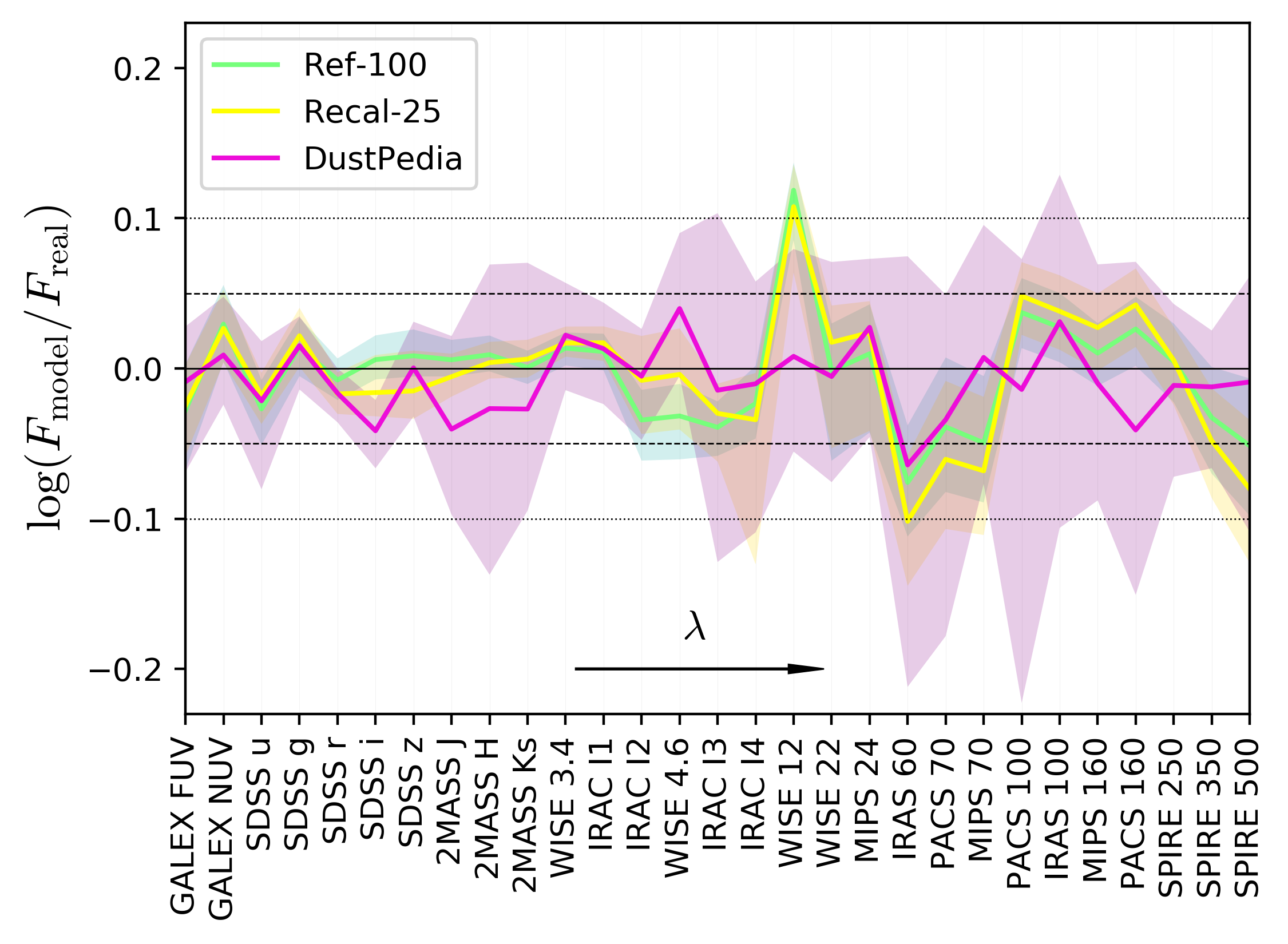}
    \caption{Ratio of the best model fluxes and the real fluxes, for the whole sample of observed and mock galaxies. Solid lines represent the median value for each band and the shaded regions show the $16\%-84\%$ range.}
    \label{fig:bestreal}
\end{figure}

We present an example of a galaxy taken from each of the three samples (Ref-100, Recal-25 and DustPedia) with their respective images\footnote{These are obtained from the public EAGLE database and the Sloan Digital Sky Survey.}, fluxes and fitted SEDs in Fig. \ref{fig:ECsed}. For all three galaxies, the model SED provides a good match to the data. This is captured in Fig. \ref{fig:bestreal}, where we show the median deviation between the observed/mock fluxes and the fitted \textsc{cigale} fluxes for each of the three galaxy samples. The highest deviation for the samples occurs in the IRAS \um{60} band (0.05-0.1 dex) and, only for the EAGLE galaxies, in the WISE \um{12} band (\char`\~0.1 dex). However, most of the model bands deviate by less than 0.05 dex from the data. As an additional test of the robustness of the fit, we performed the mock analysis described in Appendix \ref{appendix:app1}. As follows from these tests, most of the physical properties are well constrained.

\begin{figure}
	\includegraphics[width=\columnwidth]{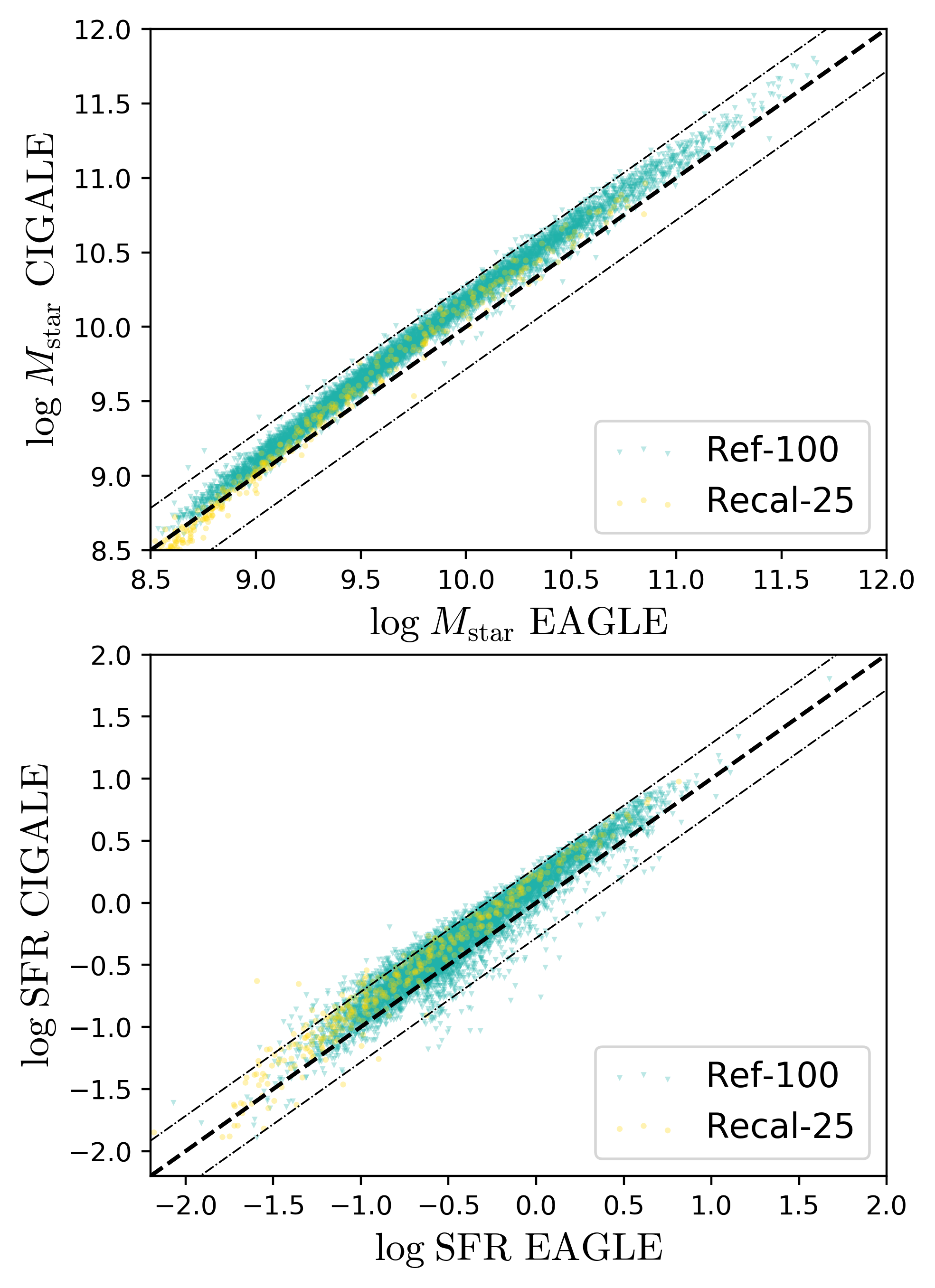}
    \caption{Comparison between the property values derived using \textsc{cigale} and the intrinsic simulation values. Dashed line represents one-to-one relation while the dash-dotted lines indicates the $\pm 0.2$ dex deviation.}
    \label{fig:app2}
\end{figure}

To validate the \textsc{cigale} method, we plot the relation between values of different properties obtained using \textsc{cigale} and the intrinsic EAGLE simulation values. Results are presented in Fig. \ref{fig:app2}. A constant offset is present, however, most of the data do not deviate more than 0.2 dex. The main cause of the difference are the assumptions we made about the star formation histories and the use of different IMFs.

All physical properties presented in this paper, unless otherwise stated, are derived from the \textsc{cigale} fits.

\section{Results}
\label{sec:results}

\subsection{MIR and FIR luminosities and colours}
\label{sec:llplots}
\begin{figure*}
 	\includegraphics[width=2\columnwidth]{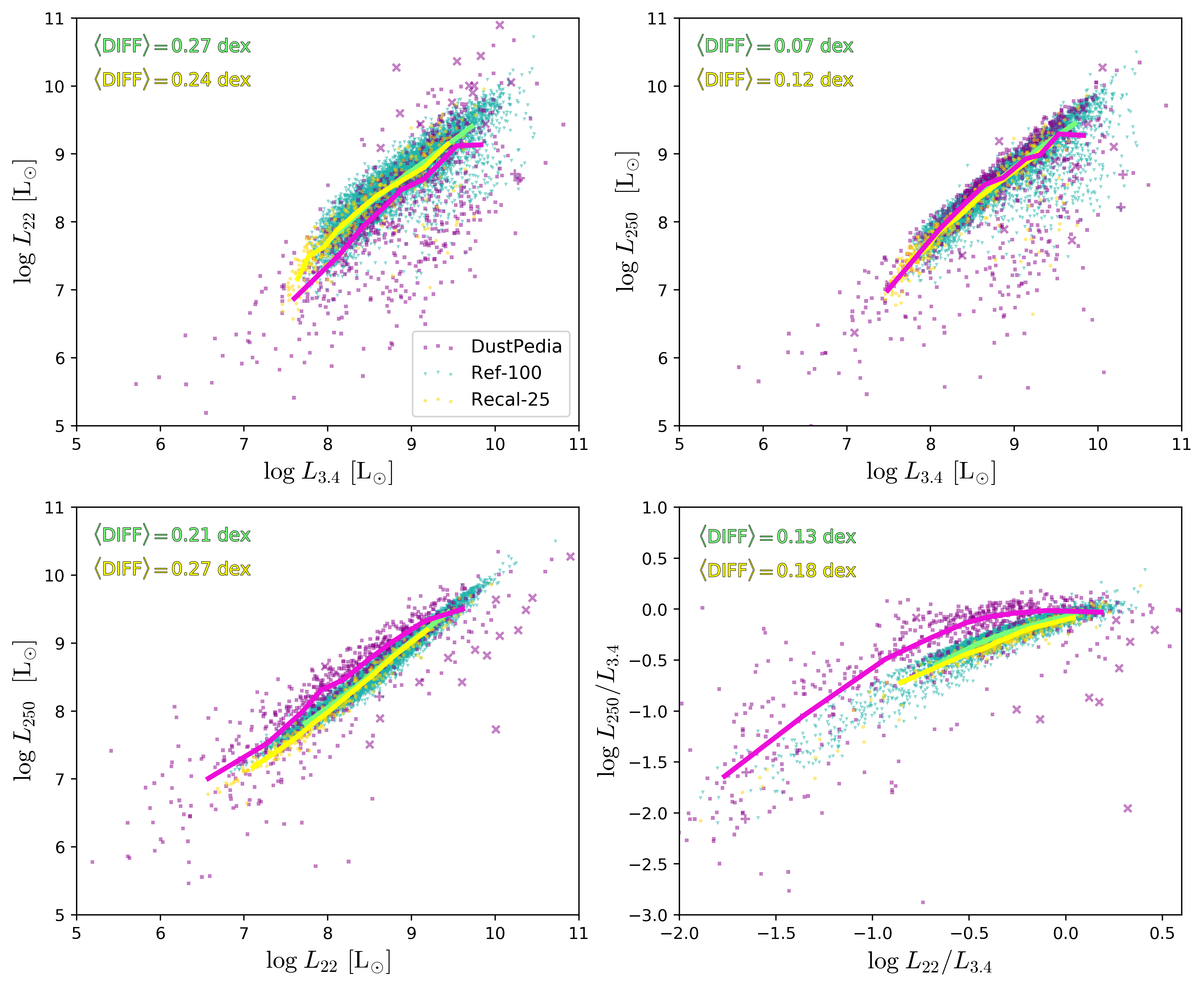}
    \caption{Relations between: (Top left) WISE \um{22} and WISE \um{3.4} luminosities; (Top right) SPIRE \um{250} and  WISE \um{3.4} luminosities; (Bottom left)  SPIRE \um{250} and WISE \um{22}  luminosities; (Bottom right) The ratios of SPIRE \um{250} and  WISE \um{22} with WISE \um{3.4} luminosity. The coloured lines represent running median. 
    The numbers in green (yellow) indicate the average offset between the running medians for Ref-100 and DustPedia (Recal-25 and DustPedia). Galaxies with an AGN (strong jet) are marked with a "$\times$" ("+").}
    \label{fig:LLLL}
\end{figure*}

We start first with an inspection of the fluxes from the DustPedia and the EAGLE databases, prior to the SED fitting. The four scatter plots of Fig. \ref{fig:LLLL} already highlight the reassuring agreement between the real and mock fluxes in the NIR to submillimetre regime. The relations for the EAGLE galaxies appear to have less scatter, than those for the DustPedia galaxies, which is expected since the EAGLE galaxies are modelled to all have the same dust properties and no observational limitations (see Sect. \ref{sec:SKIRT}). 

The top left panel of Fig. \ref{fig:LLLL} shows the relation between the WISE \um{22} and WISE \um{3.4} luminosity. As the WISE \um{3.4} band is a good proxy for the stellar mass \citep{Wen2013} and WISE \um{22} band for the SFR \citep{Lee2013}, this relation is a proxy of the main sequence of star-forming galaxies \citep{Noeske2007}. 
The main sequence of the EAGLE galaxies has already been widely investigated. \citet{Schaye2015}, \cite{Furlong2015} and \citet{Katsianis2016} compared the simulations with observations by \citet{Bauer2013} within the GAMA survey.
They all find that while the simulations typically produce specific SFRs (sSFRs) around 0.2 dex below the observational relation for the star-forming galaxies, the agreement is within the errors, with the best agreement for $M_{\mathrm{star}}>10^{10}~\mathrm{M_\odot}$ and for the high-resolution simulations. In our study, in general, all three samples follow the same trend, but with the EAGLE galaxies above the DustPedia relation.
The dearth of EAGLE galaxies at lower luminosities is a consequence of the stellar mass threshold, as discussed in Sect. \ref{sec:SKIRT}. A bimodality, representing blue and red galaxies, is reproduced: the cloud with the higher $L_{22}$ for the same $L_{3.4}$ (higher SFR for the same stellar mass) represents blue galaxies and the lower $L_{22}$ cloud red galaxies. We label a galaxy as blue or red based on its position on a $u-r$ vs. $r-z$ colour-colour diagram, with the cut-off values from \citet{Chang2015}. The EAGLE Ref-100 (Recal-25) sample contains $92\%$ ($95\%$) blue galaxies while DustPedia contains $50\%$, which explains why the bimodality is not so prominent for the EAGLE samples. Also, the running medians show an offset ($\sim 0.3$ dex) between EAGLE and DustPedia suggesting that EAGLE galaxies have either high \um{22} or low stellar mass ($L_{3.4}$) relative to the observed sample.

The top right panel represents the relation between dust and stellar mass proxies (SPIRE \um{250} and WISE \um{3.4} luminosities respectively). A strong correspondence between both the EAGLE and the DustPedia relations is found, suggesting that the offset in the top left plot is primarily driven by the $L_{22}$ output by the EAGLE-\textsc{skirt} post-processing being too high. Additionally, the bimodal distribution of the DustPedia galaxies is well reproduced by the simulations, revealing that the blue, dusty galaxies form a sequence and the red, dust-poor galaxies are located in the cloud below it. 

The bottom left panel shows the relationship between SPIRE \um{250} and WISE \um{22} luminosities. Due to our selection bias, EAGLE galaxies with little dust and low SFR are absent. The slope of the median for DustPedia is flatter for the higher $L_{22}$, which slightly changes if galaxies with AGNs are removed. 
Although the relation is fairly tight, again EAGLE galaxies show a discrepancy, having higher $L_{22}$ for the same $L_{250}$. This is also clearly visible in the bottom right panel of Fig. \ref{fig:LLLL}, where we show the relation between tracers of the specific dust mass and sSFR. The break in the median trend at $\log L_{22}/L_{3.4}\approx -0.7$ for the EAGLE samples illustrates the lack of the EAGLE galaxies with low SFR and/or high stellar mass. 

In summary, at the limited number of wavelengths we study in this section, the relations between EAGLE luminosities derived from \textsc{skirt} broadly reproduce observations. The discrepancies  are mostly coming from high $L_{22}$ for both EAGLE samples. To understand the origin of this deviation we continue our analysis in more depth in next sections. 

\subsection{Spectral energy distributions}
\label{sec:SEDs}
Complete FUV to submillimetre SEDs for all three samples (Ref-100, Recal-25, DustPedia) are extracted from the \textsc{cigale} best-model fits. First, we compare the SEDs, normalised by their bolometric luminosity and averaged, to gain insight into the sample properties. 

\begin{figure*}
	\includegraphics[width=2\columnwidth]{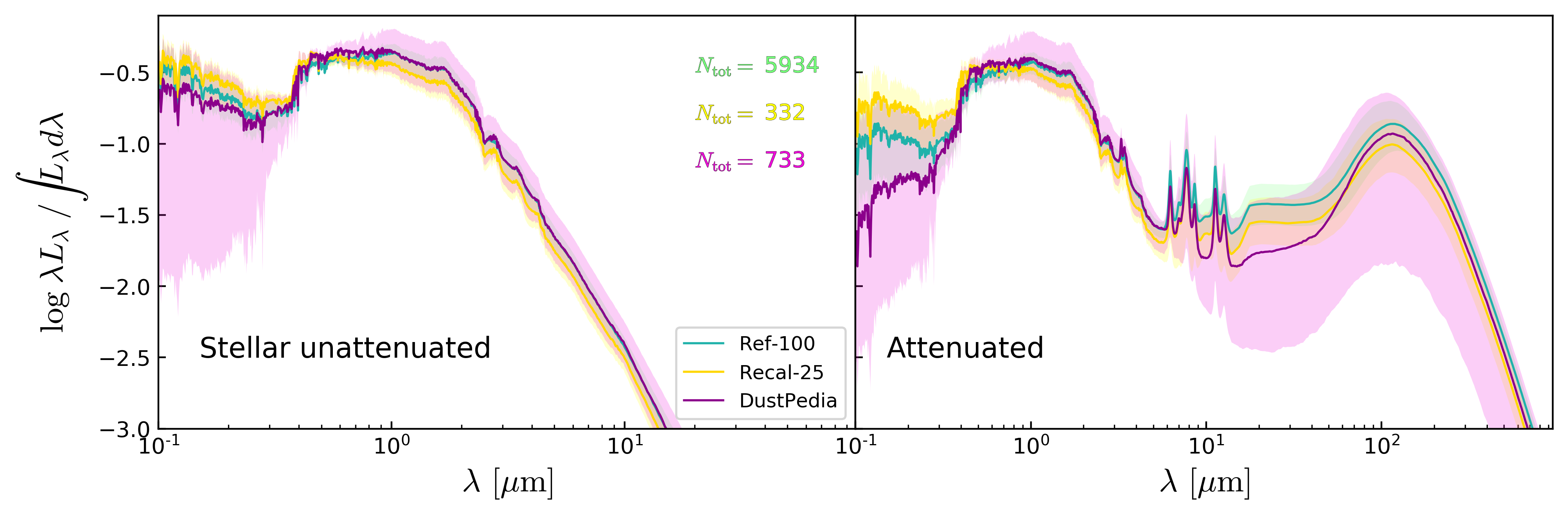}
    \caption{Median unattenuated (left panel) and attenuated (right panel) SEDs of each data-set. The plots are reconstructed from the \textsc{cigale} best fits. $10\%$ of the galaxies with the most irregular SEDs are removed from each sample. $N_{\mathrm{tot}}$ represents the actual number of galaxies used to derive these figures. The shaded regions represent $16\%-84\%$ range. }
    \label{fig:seds_sub}
\end{figure*}

Figure \ref{fig:seds_sub} shows the median SEDs\footnote{$10\%$ of the most irregular SEDs are excluded from all samples, following \citet{Bianchi2018} and \citet{Nersesian2019}.} and the regions between the 16th and 84th percentiles. Both the stellar
unattenuated (left) and the stellar attenuated (right) plots are derived from the \textsc{cigale} best fits. The left panel represents the intrinsic stellar radiation (if there were no dust). The shape of the median SEDs between the samples is similar, although the DustPedia sample shows a wider variety of SEDs compared to the EAGLE sample, especially at UV wavelengths. 
Also, the galaxies from the EAGLE samples are intrinsically slightly bluer, i.e. they emit slightly more UV and slightly less near IR (NIR) radiation compared to the DustPedia galaxies. 

The right panel shows radiation re-processed and re-emitted by dust. Compared to the intrinsic SEDs, the differences between the attenuated SEDs are much more prominent. The shape is similar in the optical and submillimetre region, but the overall spread indicates that the DustPedia sample is more diverse than both the EAGLE samples. 
This discrepancy is expected to some extent, since all EAGLE galaxies are modelled to have the same optical and calorimetric dust properties and the same dust-to-metal ratio. Additionally, the EAGLE galaxies have much less UV attenuation and more MIR radiation (see also \citealt{Baes2019}, their Fig. 1). 
In the FUV band the difference between DustPedia and Ref-100 (Recal-25) is 0.37 dex (0.59 dex). 
The difference in the WISE \um{22} band is slightly lower: 0.34 dex (0.21 dex). We shall return to these differences in the SEDs in Sect. \ref{sec:pop}. 

\subsection{Physical properties and dust scaling relations}
\label{sec:proxies} 
In this section, we examine how well the physical properties, derived using \textsc{cigale}, are represented by their common proxies, and then we focus on the scaling relations between different physical properties.

\begin{figure}
	\includegraphics[width=\columnwidth]{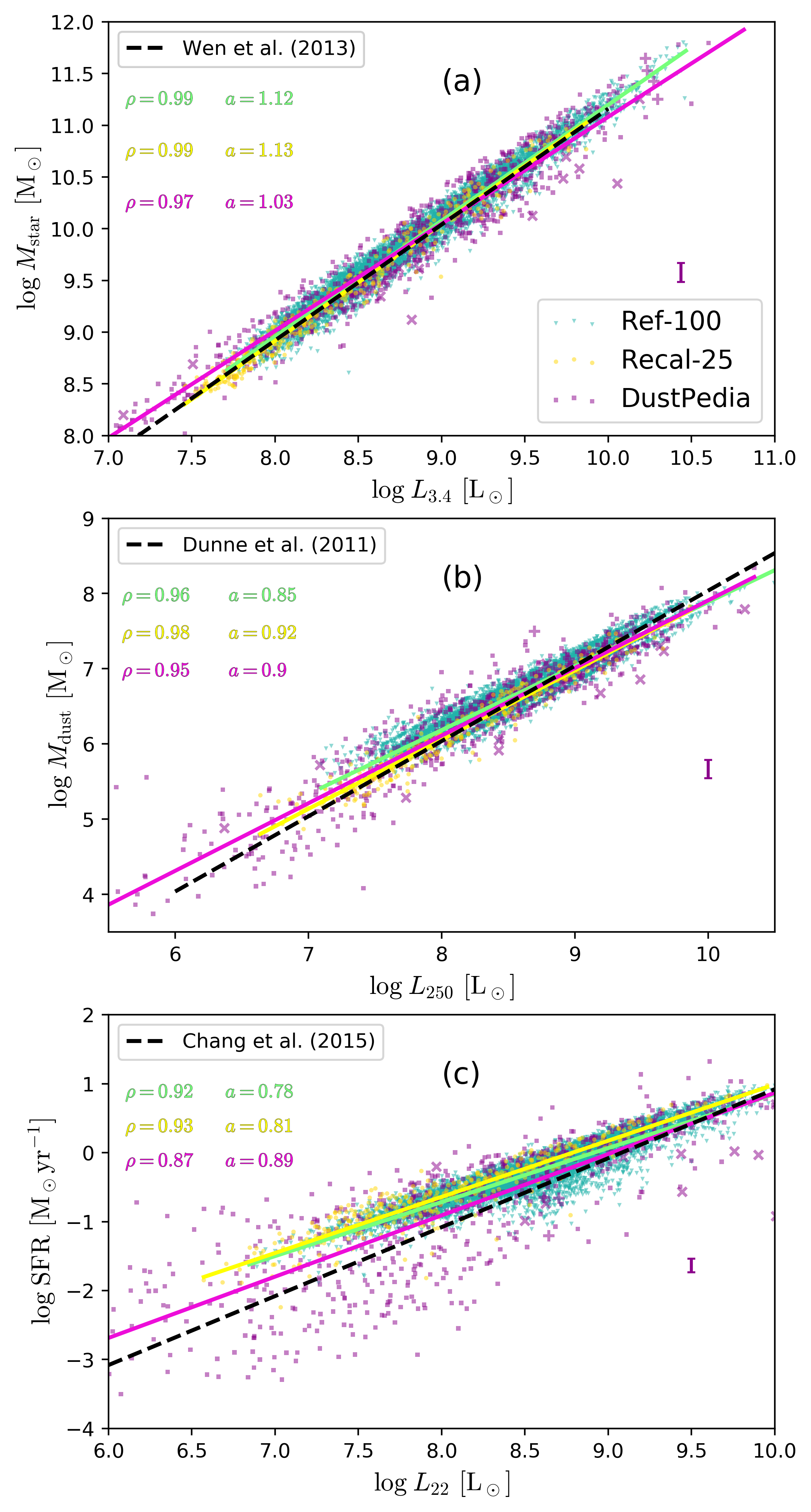}
    \caption{Luminosity proxies and appropriate properties. (a) Stellar mass versus WISE \um{3.4} band, (b) dust mass versus SPIRE \um{250} band, and (c) SFR versus WISE \um{22} band. All physical properties are inferred from \textsc{cigale}. Coloured lines represent linear fits and their length indicates the sample domain. The dashed black line is the fit from the literature. The error bar in each panel represents the median error for the DustPedia sample. Spearman coefficients $\rho$, and slopes $a$ for each relation and data-set are indicated as well. Galaxies with an AGN (strong jet) are marked with a "$\times$" ("+").}
    \label{fig:proxies}
\end{figure}

Figure \ref{fig:proxies}a demonstrates the quality of the WISE \um{3.4} luminosity as a stellar mass tracer, since it is sensitive to the evolved stellar populations that dominate the baryonic mass in galaxies, and at the same time it is not very affected by the dust attenuation \citep{Norris2014}. Solid lines represent linear fits to the data while the dashed line is the fit from \citet{Wen2013}. The three samples agree very well, part from the low stellar mass end which is below the EAGLE mass thresholds. The Spearman rank-order correlation analysis coefficient (indicated in the plots) is slightly higher for the EAGLE sample than for the DustPedia sample.  
 
Considering that most of the dust is in the cold phase, submillimetre radiation can be used as a proxy for the dust mass in galaxies \citep{Dunne2011, Eales2012}. Fig. \ref{fig:proxies}b shows the level of agreement of our samples. 
The highest Spearman coefficient is for Recal-25. The linear fits agree with those of \citet{Dunne2011}. The lack of simulated galaxies with low SPIRE \um{250} luminosities is, again, caused by our chosen selection of EAGLE galaxies (see Sect. \ref{sec:SKIRT}). 

Fig. \ref{fig:proxies}c shows the relation between the WISE \um{22} luminosity and the SFR. From the abundance of different SFR tracers \citep[e.g.][]{Hao2011, Cortese2012, Lee2013, Boquien2016, Casasola2017}, in this study, we are focusing on a reliable single MIR band SFR tracer \citep{Calzetti2010, Cluver2017}. Fig. \ref{fig:proxies}c  highlights the differences between the EAGLE and the DustPedia samples. 
The lower Spearman coefficient of the DustPedia sample is, together with the slope and the scatter of the relation, primarily driven by the large number of galaxies with low SFRs and luminosities at \um{22}. 
In the domain below $10^8~\mathrm{L_\odot}$, the spread in SFR is almost two orders of magnitude, at fixed \um{22} luminosity for the DustPedia sample, whereas the EAGLE galaxies continue to lie along a fairly well-defined sequence. 
We revisit this analysis in Sect. \ref{sec:diss}, where we split each sample based on the galaxy sSFR.

\begin{figure*}
	\includegraphics[width=2\columnwidth]{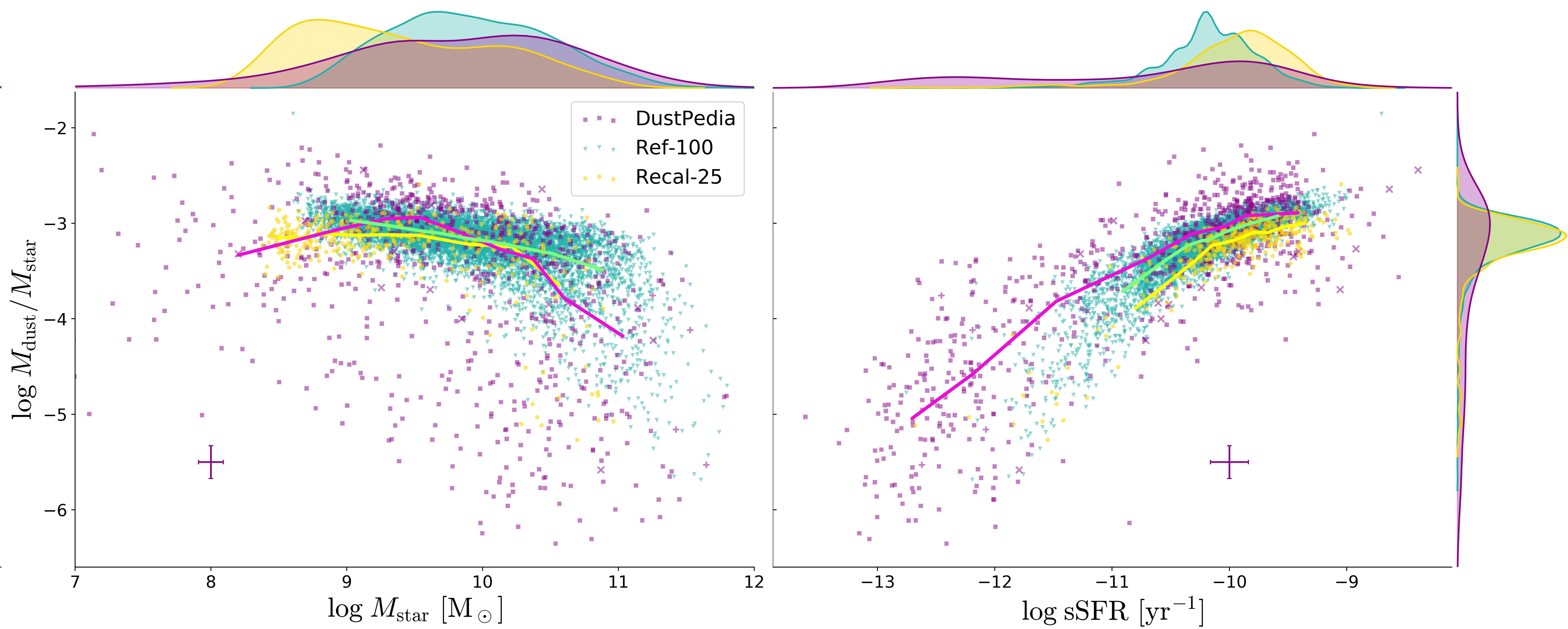}
    \caption{Scatter plots represent relations between dust-to-stellar mass ratio versus stellar mass (left) and sSFR (right). All properties are inferred from \textsc{cigale}. On the top and right normalised distributions of each property are shown. The coloured curves indicate running median. The error bars correspond to the medians of the errors for the DustPedia sample. Galaxies with an AGN (strong jet) are marked with a "$\times$" ("+").}
    \label{fig:scal}
\end{figure*}

In Fig. \ref{fig:scal}, we consider the same and the analogous  scaling relations as those of \citet{Camps2016} used to calibrate the free parameters in the post-processing procedure (see Sect. \ref{sec:SKIRT}): $M_{\mathrm{dust}}/M_{\mathrm{star}}$ versus stellar mass, and $M_{\mathrm{dust}}/M_{\mathrm{star}}$ versus sSFR (instead of $\mathrm{NUV}-r$ colour). We decided to compare physical properties, since they are derived in the same way, and  $\mathrm{NUV}-r$ colour is generally assumed to be a proxy for sSFR \citep{Salim2005,Salim2007, Schiminovich2007}. 
We investigate whether these scaling relations are still valid considering that the samples of both observed and simulated galaxies are now larger, that all properties are derived in a self-consistent way and the post-processing procedure on the EAGLE galaxies is slightly modified, as explained in Sect. \ref{sec:SKIRT}.

The left panel of Fig. \ref{fig:scal} shows the relation between specific dust mass and stellar mass. The figure indicates overall agreement, although some discrepancies are present. First, the large scatter found for the DustPedia sample is absent for both EAGLE samples. 
The reason is twofold. Firstly, observational limitations are not accounted for. Secondly, the majority of the scattered DustPedia galaxies are either low stellar mass galaxies with high sSFR that are too metal-poor to have formed much dust \citep{DeVis2019}, or early-types with very little dust. Both  populations are missing in the EAGLE samples due to the stellar and the dust mass thresholds, respectively. Additionally, a companion DustPedia observational paper \citep{Casasola2020} studies the same relation focused on the late-type galaxies, showing indeed, less dispersion. 
The relation for EAGLE is flatter than that of the DustPedia galaxies, as also noticed by \citet{Camps2016}, comparing to the HRS sample. The median dust-to-stellar mass ratio of the high resolution Recal-25 run is systematically lower than that of Ref-100 (average difference in the overlapping bins is 0.1 dex). This is expected since Recal-25 has a lower dust detection limit, allowing for less dusty galaxies to make the selection criterion.

The right panel of Fig. \ref{fig:scal} represents a relation between the specific dust mass and the sSFR. This relation is analogous to the one in the bottom right panel of Fig. \ref{fig:LLLL}. 
Here we also have DustPedia galaxies evenly distributed over a large sSFR range, while the EAGLE galaxies, due to our selection effect, are mostly clustered in the high sSFR region. 
Galaxies from the Recal-25 sample on average have the highest sSFR (0.2 dex more than Ref-100, and 0.4 dex more than DustPedia in the overlapping bins). 
The difference in the sSFR between the EAGLE samples can be a consequence of a higher intrinsic sSFR of Recal-25, noticed in previous studies \citep[e.g.][]{Schaye2015,Furlong2015}.
At the same time, these studies also find that the EAGLE sSFRs generally tend to slightly underestimate observations, contrary to our results (see also Fig. \ref{fig:LLLL}, top left panel).
However, they concentrate on the star-forming galaxies and our approach has an important advantage of using the same methods of deriving physical properties and their fits, for all three samples.

Additional discrepancy is that the median line of the DustPedia sample in the right plot is systematically higher than for the EAGLE samples, which is not observed in the \citet{Camps2016} study with the $\mathrm{NUV}-r$ colour. This inconsistency can be caused by the difference in methods of acquiring properties between this study and that of \citet{Camps2016}. We expect more accurate results with our method, which uses the complete galaxy SED, while they incorporated only the limited number of bands.

Another important property of a galaxy is the amount of energy absorbed by dust. It is defined as the ratio between the dust luminosity and the bolometric luminosity: 
 $$f_{\mathrm{abs}}=\frac{L_{\mathrm{dust}}}{L_{\mathrm{bolo}}}$$ This ratio contains information on how optically thick a galaxy is, which further depends on the amount, composition, and geometry of dust in a galaxy. 
Previously, studies reported the average value of $f_{\mathrm{abs}}$ to be $0.25\pm 0.05$ with the highest value for late-type galaxies \citep[and references therein]{Davies2012, Viaene2016, Bianchi2018}. 

\begin{figure}
	\includegraphics[width=\columnwidth]{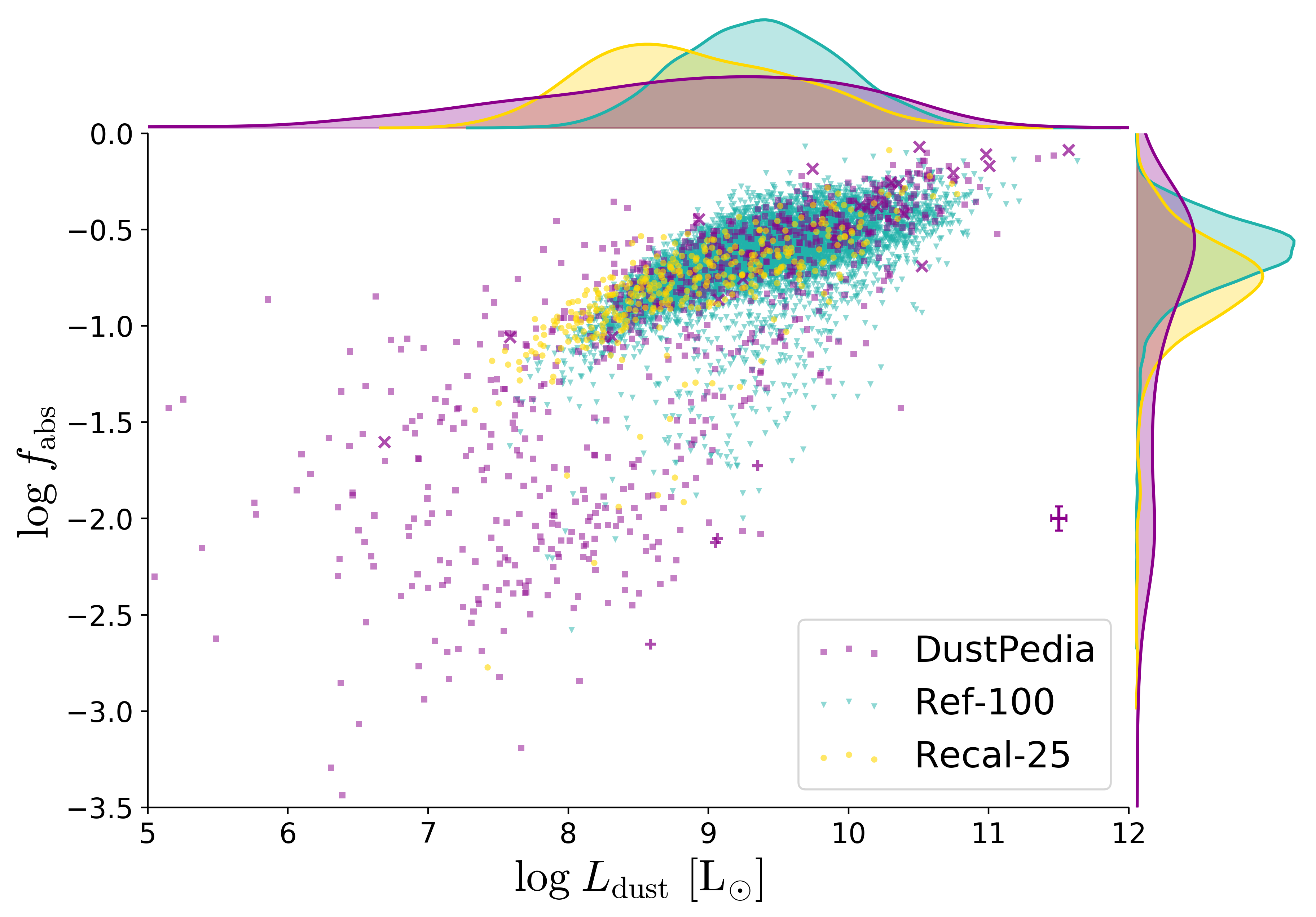}
    \caption{Amount of energy absorbed by dust versus dust luminosity, as inferred from \textsc{cigale}. The error bar corresponds to the median of the errors for the DustPedia sample. Galaxies with an AGN (strong jet) are marked with a "$\times$" ("+").}
    \label{fig:fabs}
\end{figure}

The most extensive study of $f_{\mathrm{abs}}$ is by \citet{Bianchi2018}, where they used \textsc{cigale} to investigate the extent to which dust affect the stellar light in the DustPedia sample, and its correlation with different galaxy properties. 
They find weak trends with morphological type and some physical properties (e.g. $M_{\mathrm{star}}$ and sSFR), and a moderate correlation with bolometric and dust luminosity for the late-type sub-sample. 
We repeated the same methodology for the EAGLE galaxies. The results of this consistent comparison are shown in Fig. \ref{fig:fabs}, where we compare $f_{\mathrm{abs}}$ with dust luminosity. 
Most of the EAGLE galaxies occupy the higher $f_{\mathrm{abs}}$ area, indicating that most of the simulated galaxy sample has enough dust to reprocess notable amounts of stellar light. 
The DustPedia galaxies with the highest $f_{\mathrm{abs}}$ are those with the AGN flag, however, not all of them have $f_{\mathrm{abs}}$ that high \citep{Bianchi2018}. All three samples show two streams that connect around $\log L_{\mathrm{dust}} [\mathrm{L_{\odot}}]=10 $ and  $\log f_{\mathrm{abs}}=-1$. 
In Sect. \ref{sec:pop} we will tackle in detail the differences in the scatter.
 
\section{Discussion}
\label{sec:diss}

The results from this study so far indicate overall agreement between simulations and observations, apart from a few discrepancies, mostly differences in the scatter and offsets in the relations and the SED regimes associated with SFR. This confirms earlier findings of \citet{Baes2019}, who reported small but systematic tensions in certain sections of the cosmic SED. In general, these may arise from: a different galaxy population mix in the three samples, and/or differences in the stellar/dust properties of EAGLE and DustPedia, coming from the imperfections of the post-processing procedure or limitations in EAGLE recipes for galaxy formation. In the following sections, we will address each of these possible causes for deviations.

\subsection{The galaxy population}
\label{sec:pop} 
\begin{figure*}
	\includegraphics[width=2\columnwidth]{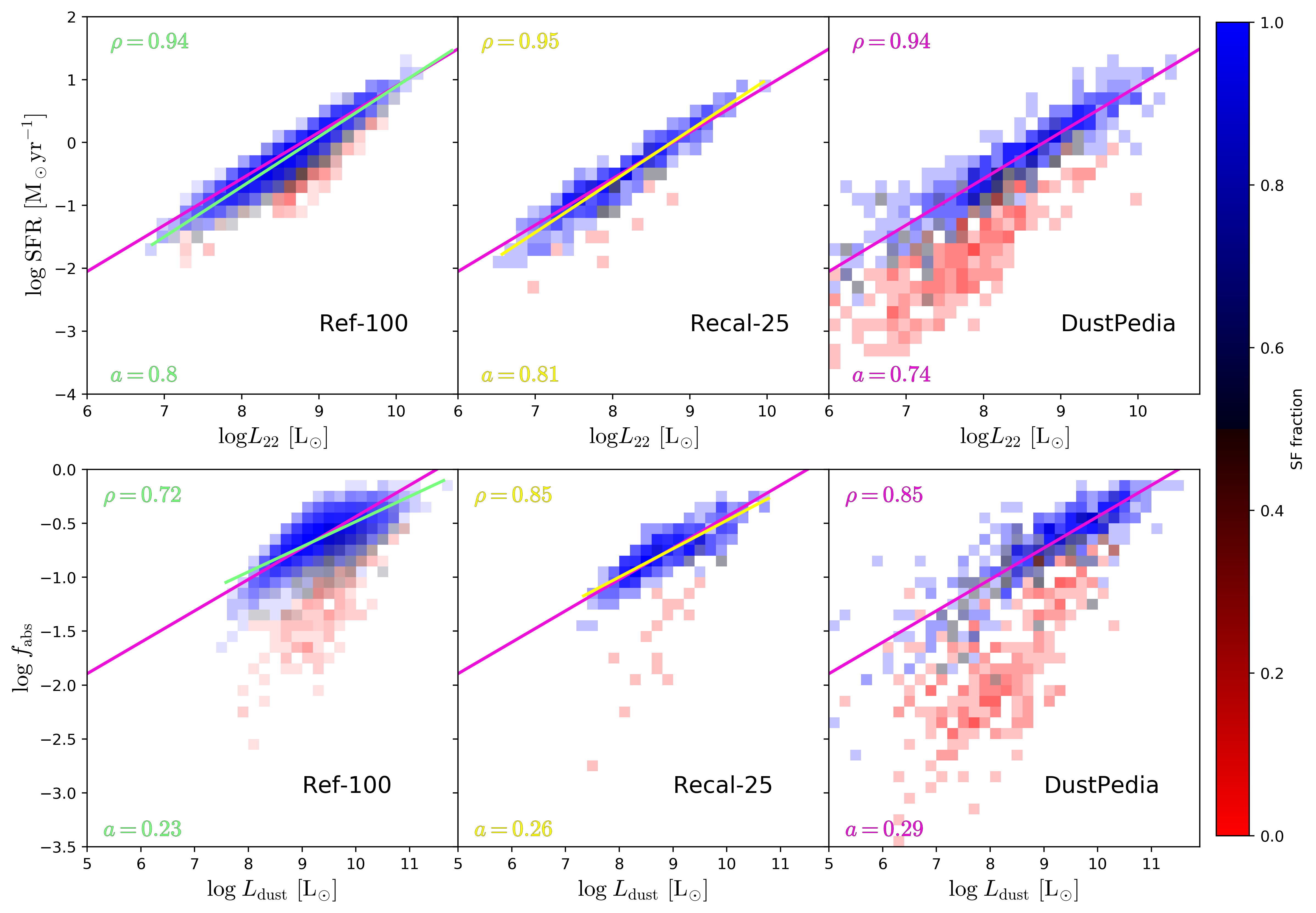}
    \caption{Top: Same as Fig. \ref{fig:proxies}c with each sample in the different panel. Bins are colour-coded by the fraction of the star-forming galaxies in the bin. Magenta line represents a fit only through star-forming DustPedia galaxies. Green and yellow lines indicate fits of the EAGLE run in the appropriate panel, again only for the star-forming galaxies. Spearman coefficients $\rho$, and slopes $a$ for each data-set are indicated in the top left corner. Transparency indicates the number of galaxies in each bin. Bottom: Same as Fig. \ref{fig:fabs} with each sample in the different panel, analogue to the top row.}
    \label{fig:SFRL22_fabs}
\end{figure*}

As discussed in Sect. \ref{sec:proxies}, relations between different galaxy properties and appropriate luminosity proxies for all three samples are reasonably tight and in agreement, except the relation between WISE \um{22} luminosity and SFR, where the discrepancies between the three samples are higher (see Fig. \ref{fig:proxies}c). 
To better understand the differences between the three galaxy samples, we further analyse this relation. 
The top row of Fig. \ref{fig:SFRL22_fabs} represents the same as Fig. \ref{fig:proxies}c with the different samples in the different panels. 
the colour-coding is based on the fraction of star-forming galaxies in each bin. 
We assume a galaxy is star-forming if its sSFR satisfies $\log \mathrm{sSFR}>-10.8 ~\mathrm{yr^{-1}}$ \citep{Salim2014}. 
This figure clearly indicates that the discrepancy in the $\mathrm{SFR} - L_{22}$ relation between the three samples is primarily due to a different galaxy mix in the samples. 
For all three, star-forming galaxies form the same tight sequence. 
The main distinction is the large fraction of galaxies in the DustPedia sample with WISE \um{22} radiation that does not solely relates to the star formation activity, but that is mostly arising from the evolved stars or the warm dust heated by them \citep{Madden1999,Xilouris2004, Simonian2017}. 
These quiescent galaxies are removed from the EAGLE-\textsc{skirt} sample, since they lack sufficient amounts of dust to make our selection criterion. 
Accordingly, when we consider only star-forming galaxies (i.e. galaxies in the blue bins), and compare to Fig.\ref{fig:proxies}c, the scatter is largely reduced: the Spearman coefficients are now almost the same for all three samples, and the gradient of the power-law fit for DustPedia is lower than the one for the full sample, agreeing better with the simulations. 

In Sect. \ref{sec:proxies} we demonstrated that the global dust emission properties, represented by the $f_{\mathrm{abs}}$ versus  $L_{\mathrm{dust}}$ relation, agree for the three samples, apart from the region of low dust emission (see Fig. \ref{fig:fabs}). 
Now, following the same approach as for the $\mathrm{SFR}-L_{\mathrm{22}}$ relation, we inspect if the differences in the galaxy populations are again driving the tension between the three samples. 
In the bottom row of Fig. \ref{fig:SFRL22_fabs}, we compare $f_{\mathrm{abs}}$ with dust luminosity. 
The samples are in the different panels, colour-coded by fraction of star-forming galaxies in each bin. 
It is apparent from the figure that the differences in the galaxy population mix affect the relation notably: when we neglect the dust-poor galaxies, the results become comparable. For instance, although Recal-25 has a lower average value of $f_{\mathrm{abs}}$ (as already seen from the distribution in Fig. \ref{fig:fabs} for the whole sample), star-forming galaxies from this sample follow a hardly distinguishable trend from the one of DustPedia. 
In comparison, a slight offset is seen for the Ref-100 sample, which is expected considering the Recal-25 run has a higher resolution and better sampled disks \citep{Trayford2017}. 
For a galaxy property that depends on the geometry of a galaxy, such as $f_{\mathrm{abs}}$ \citep{Viaene2016, Bianchi2018}, a sample with more realistic spiral galaxies will reach a better agreement with observations.

\begin{figure*}
	\includegraphics[width=2\columnwidth]{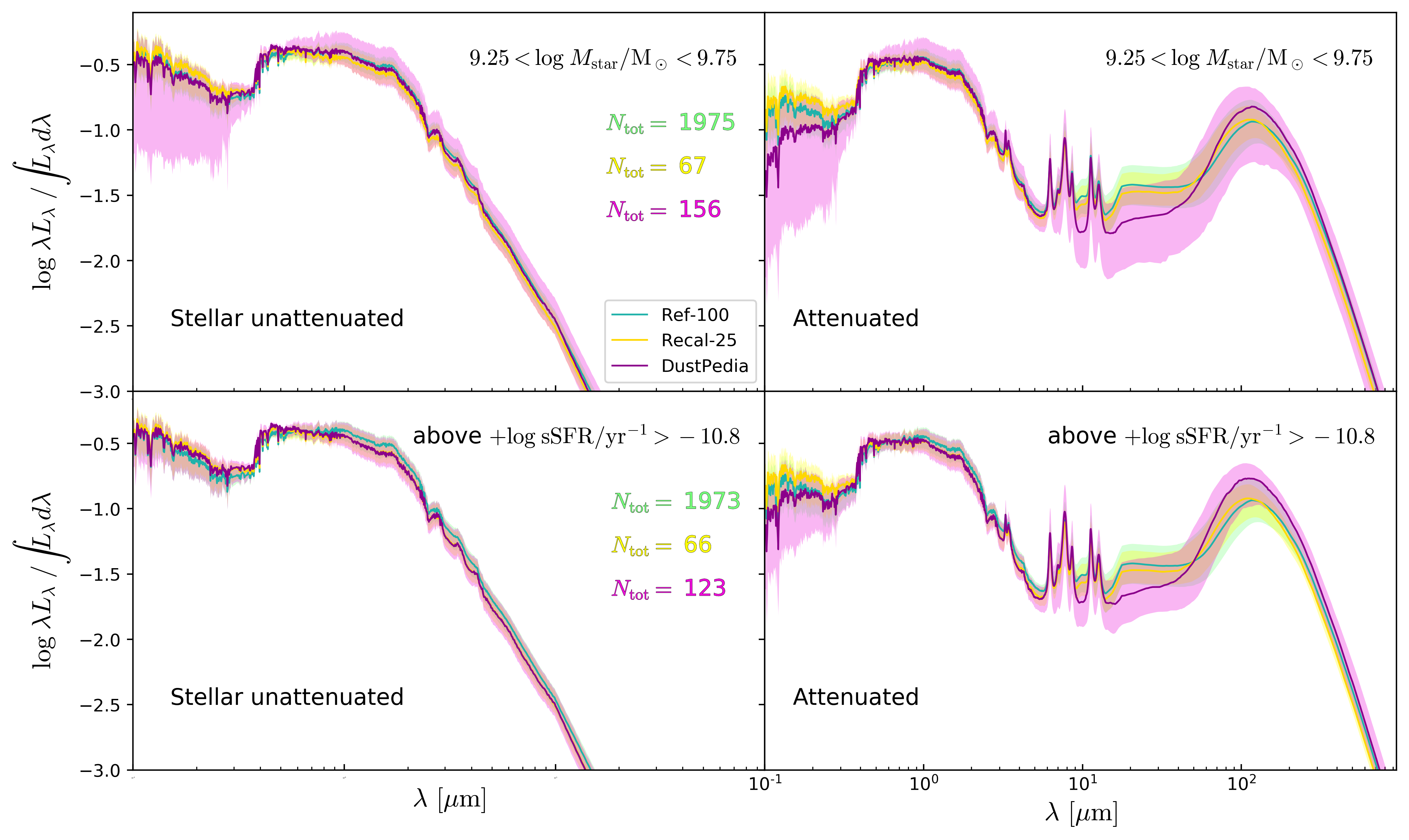}
    \caption{Similar to Fig. \ref{fig:seds_sub} with the top row including only galaxies in the indicated stellar mass range while the bottom row excludes also the passive galaxies.}
    \label{fig:SEDs_bins}
\end{figure*}

Based on Fig. \ref{fig:SFRL22_fabs}, we are inclined to conclude that differences in the galaxy population between the samples can explain the tensions in the relations discussed in Sect. \ref{sec:results}. 
However, these differences alone can not explain the substantial offset between EAGLE and DustPedia along the whole range of $L_{22}/L_{3.4}$ luminosity ratios seen in Fig. \ref{fig:LLLL} bottom right. 
To understand how strong the effect of the different sample selection actually is, we examine the median SEDs again. 
We select galaxies in the small mass range of $9.25<\log M_{\mathrm{star}}/\mathrm{M_{\odot}}<9.75$ (based on the \textsc{cigale} results), and reanalyse the features of their SEDs. 
Figure \ref{fig:SEDs_bins} is analogous to Fig. \ref{fig:seds_sub}, with the top row including only galaxies in the selected stellar mass range. 
In this range, most of the galaxies are late-type with a significant amount of dust, which corresponds better to the EAGLE sample. 
There is a clearly better agreement between DustPedia and EAGLE, but the discrepancies remain in both the UV and MIR part of the galaxy spectrum. 
In the attenuated FUV band, the median DustPedia-Ref-100 (DustPedia-Recal-25) difference is 0.19 dex (0.27 dex). 
For the WISE \um{22} band, the difference is 0.25 dex (0.21 dex, same as when comparing full samples). In later sections we discuss discrepancies in these bands for a range of mass bins.

For a fixed $M_{\mathrm{star}}$, the the DustPedia sample contains both star-forming and passive early-type galaxies (in this bin, $\approx 21\%$ are early-type galaxies). 
For a direct comparison with the EAGLE sample, we thus also included an additional constraint: $\log \mathrm{sSFR}>-10.8 ~\mathrm{yr^{-1}}$, since galaxies with a lower sSFR are mostly passive. 
The new constraint minimally affects the EAGLE-SKIRT sample, because almost all galaxies in the selected mass bin already have the sSFRs above $-10.8 ~\mathrm{yr^{-1}}$. This is in agreement with \citet{Katsianis2019}, who found the fraction of only $\approx 0.1$ early-type galaxies at similar $M_{\mathrm{star}}$ and sSFRs, for the whole EAGLE reference simulation, at $z=0$.
The results are shown in bottom panels of Fig. \ref{fig:SEDs_bins}. 
Showing the unattenuated stellar light only, the left panel shows an almost perfect agreement. The discrepancies in the attenuated SEDs are smaller, but still present: in FUV band the difference for DustPedia-Ref-100 (DustPedia-Recal-25) is 0.08 (0.16), and in WISE \um{22} band the difference for DustPedia-Ref-100 (DustPedia-Recal-25) is 0.21 (0.16) dex. 
The average difference in the FIR seems to have increased. Here, we can expect that the differences in the dust models (by \citet{Zubko2004} in the post-processing, and THEMIS in the SED fitting) have an effect (see Sect. \ref{sec:cigale}). The prominent features around \um{20} modelled by \citet{Zubko2004}, and then fitted by THEMIS would produce an excess at these wavelengths, as the model would try to fit PAH features with the continuum emission.

The findings of this section confirm that the differences in the galaxy populations between the three samples greatly influence, yet can not fully explain the differences seen in the scaling relations and the SEDs.

\subsection{The EAGLE simulations and the \textsc{skirt} post-processing}
\label{sec:ppp}
We now turn to another possible cause for the differences between simulations and observations: the limitations of the EAGLE simulations and the procedure applied to the EAGLE particle data, in order to incorporate dust in the simulations (see Sect. \ref{sec:SKIRT}). 

A growing number of studies have analysed the EAGLE simulations and how well they reproduce different observables that were not used for their calibration \citep[e.g.][]{Schaye2015,Crain2015,Furlong2015,Lagos2015,Trayford2015,Trayford2016, Furlong2017, Katsianis2017, Tescari2018}.
Considering the good agreement on the stellar emission only (Fig. \ref{fig:SEDs_bins}, left panels), we also conclude that the stellar properties of the EAGLE galaxies are representative of the stellar properties of our sample of real galaxies.
Additionally, as explained in Sect. \ref{sec:cigale}, we performed a check where we compare the values of intrinsic properties of EAGLE with the values derived from \textsc{cigale} and we find that the data are in agreement.

In the remainder of this section we focus on the impact of our post-processing recipe on the results. During the calibration of the parameters in the post-processing, in the IR domain, only SPIRE bands were used \citep{Camps2016}. This  implies that the MIR-FIR regime of the galaxy spectrum is essentially unconstrained and susceptible to discrepancies. 
In Sect. \ref{sec:SEDs}, we have already seen that at these wavelengths the median SEDs are discrepant, even if we limit the analysis to a specific stellar mass and sSFR bin (Fig. \ref{fig:SEDs_bins}). Thus, it may be assumed that they are caused by the characteristic treatment of the star-forming regions, applied using the \textsc{mappings-iii} templates. 

To investigate this further, we analyse a galaxy scaling relation based on the UV and dust emission. 
To lessen the effect of the different galaxy populations in each sample, we concentrate only on the star-forming galaxies, i.e. galaxies with $\log \mathrm{sSFR}>-10.8 ~\mathrm{yr^{-1}}$. We analyse the $\mathrm{IRX}-\beta$ relation presented by \citet{Meurer1999}, where IRX is the infra-red excess defined as: $$\mathrm{IRX}=\log \frac{L_{\mathrm{dust}}}{L_{\mathrm{FUV}}}$$ while $\beta$ is the UV slope defined as: $$\beta=\frac{\log f_{\nu(\mathrm{NUV})}/f_{\nu(\mathrm{FUV})}}{\log \lambda_{\mathrm{NUV}}/ \lambda_{\mathrm{FUV}}}-2,$$ where $f_\nu$ is the flux density. The relation, considering it is sensitive to the dust attenuation, has been thoroughly analysed for a wide range of redshifts  \citep[e.g.][]{Meurer1999, Kong2004, Overzier2011, Boquien2012, Salim2019}. The relation demonstrates that star-bursting galaxies have redder UV (towards positive $\beta$ values) colour if more UV radiation is reprocessed by dust.

\begin{figure}
	\includegraphics[width=\columnwidth]{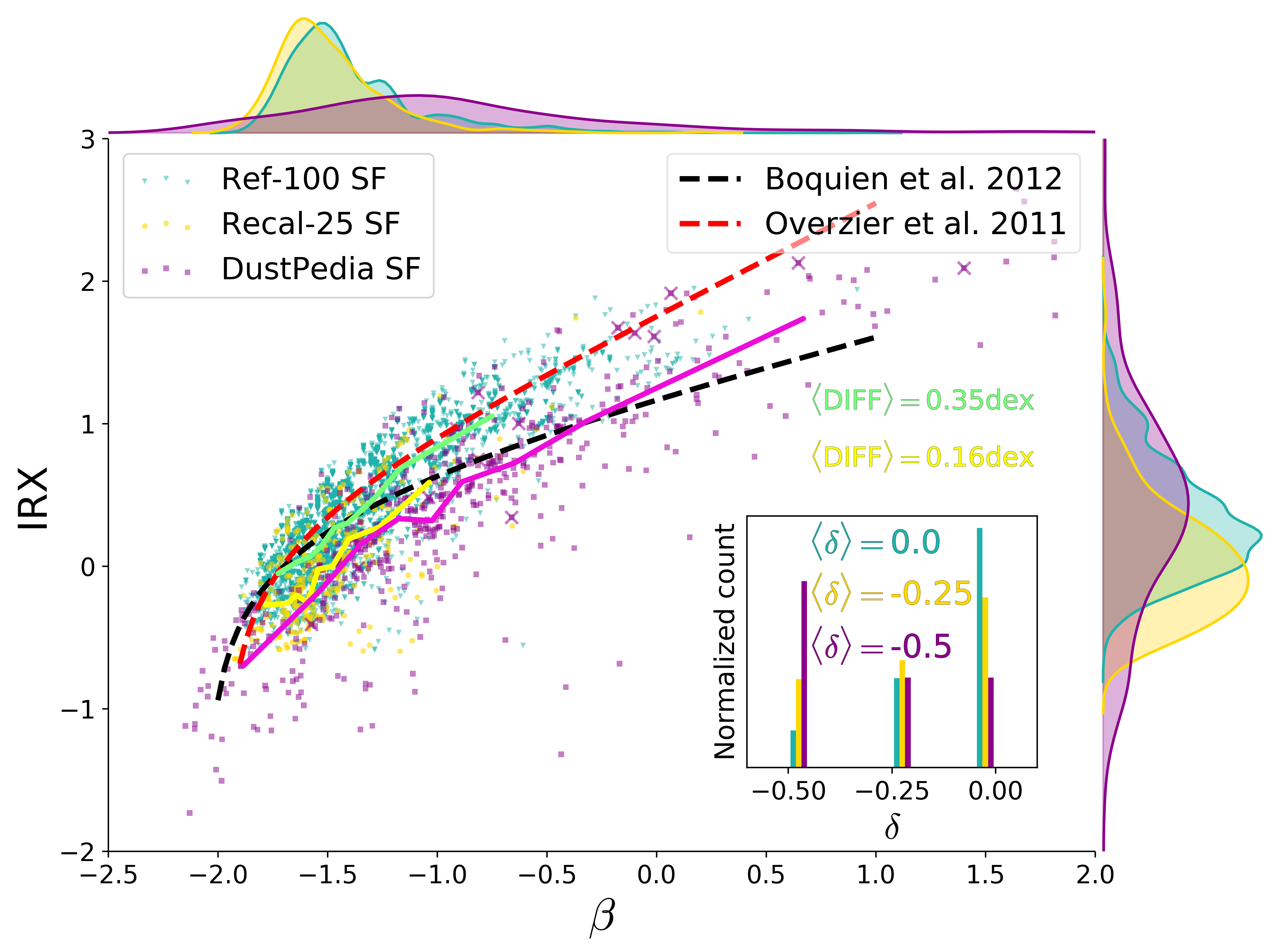}
    \caption{IR excess versus UV slope, only for the star-forming sub-samples. The dashed lines represent fits from the literature, solid lines are the running medians. The number in green (yellow) represents the average deviation of the Ref-100 (Recal-25) from the DustPedia running median. The inset shows the distribution of the attenuation curve slopes, as inferred from \textsc{cigale}.}
    \label{fig:IRXb}
\end{figure}

Figure \ref{fig:IRXb} shows the $\mathrm{IRX}-\beta$ relation for star-forming galaxies in all three samples. 
All properties needed to calculate IRX and $\beta$ are derived using \textsc{cigale}. Dashed lines are fits for star-bursting galaxies from \citet{Overzier2011} and \citet{Boquien2012}. 
Remarkably, the EAGLE SEDs fits agree well within the range of these observational relations, comparable to the previous studies that included cosmological simulations \citep[e.g.][]{Narayanan2018,Hou2019}.  
In general, star-forming EAGLE galaxies show the same trend as star-forming DustPedia galaxies. However, Ref-100 shows a small but systematic offset that can be partially attributed to resolution and the fact that Recal-25 has a higher UV output (since it has on average higher sSFR, \citealt{Schaye2015}) and hence a smaller IRX.

Another origin of the discrepancy comes from the difference in the attenuation curve. 
As mentioned in Sect. \ref{sec:cigale}, our CIGALE models include a modified \citet{Calzetti2000} attenuation curve, implemented with a free parameter slope and without a UV bump \citep{Nersesian2019}. 
The distribution of slopes for all samples is shown in the inset of Fig. \ref{fig:IRXb}, where the DustPedia sample has the steepest median slope and Ref-100 the shallowest. \citet{Salim2019} analysed $\mathrm{IRX}-\beta$ relation for around 23,000 low-redshift galaxies from the GSWLC-2 sample \citep{Salim2018} and argue that the scatter and the offset from the \citet{Overzier2011} curve are driven by the diversity of the attenuation curves, mainly their slopes. 
They demonstrate\footnote{Although using nonzero UV bump strength.} that the shallower the slope is, the higher the galaxy is in the $\mathrm{IRX}-\beta$ plot, which is reproduced in our study.   

The offset between the median lines, seen in the $\mathrm{IRX}-\beta$ relation, is caused by the offset in the median slope of the attenuation curve. \citet{Narayanan2018b} investigated the main influence on the diversity of the slopes on a sample of "zoom-in" simulated galaxies and they highlight the importance of the star-to-dust geometry, i.e. high fraction of obscured young and low fraction of obscured old stars steepen the attenuation curve. Interpreting results in this context, we argue that our sub-scale modelling of the H\textsc{ii} regions in our EAGLE-\textsc{skirt} post-processing algorithm can be improved, possibly by changing the value of $f_{\mathrm{PDR}}$. We note that \citet{Trayford2019b}, analysing the slope of the attenuation curves of the EAGLE galaxies, found that Recal-25 does not have the steepest slope, contrary to our result. However, they inspected only the diffuse dust while the effect of the birth clouds will steepen the curve for the galaxies with more star formation, since the spectrum of the young, blue stars will be greatly reddened. Recal-25 has intrinsically higher sSFR than Ref-100 which drives the slope towards steeper values. Additionally, their galaxy sample has a different stellar, dust mass and redshift thresholds, and they calculated the attenuation at a fixed dust surface density. Furthermore, to calculate the attenuation, \citet{Trayford2019b}, used two different \textsc{skirt} runs (with and without diffuse dust), while we use \textsc{cigale} derived results on one \textsc{skirt} run.

\subsection{Towards an improved post-processing recipe}
\label{sec:irxb}
Taken together, the results from this study indicate that the derivation of fluxes (in a number of bands), therefore SEDs, and consequently physical properties of the EAGLE galaxies, could be improved. In principle, we could repeat the calibration exercise by \citet{Camps2016} and search the entire  space of the subgrid parameters for the combination that minimises the tension between EAGLE and DustPedia, using the relations presented in this paper. 
However, such a wide search is a daunting task, given the number of possible free parameters in the subgrid recipe, and the computational cost to run radiative transfer simulations for the EAGLE galaxies, even at $z=0$. Instead, we can use the observed differences to guide us in which direction the post-processing calibration should be heading. 
Concretely, we investigate whether the flux differences are correlated with any particular physical properties, and whether these correlations uncover additional effects of the post-processing procedure and its calibration.

\begin{figure}
	\includegraphics[width=\columnwidth]{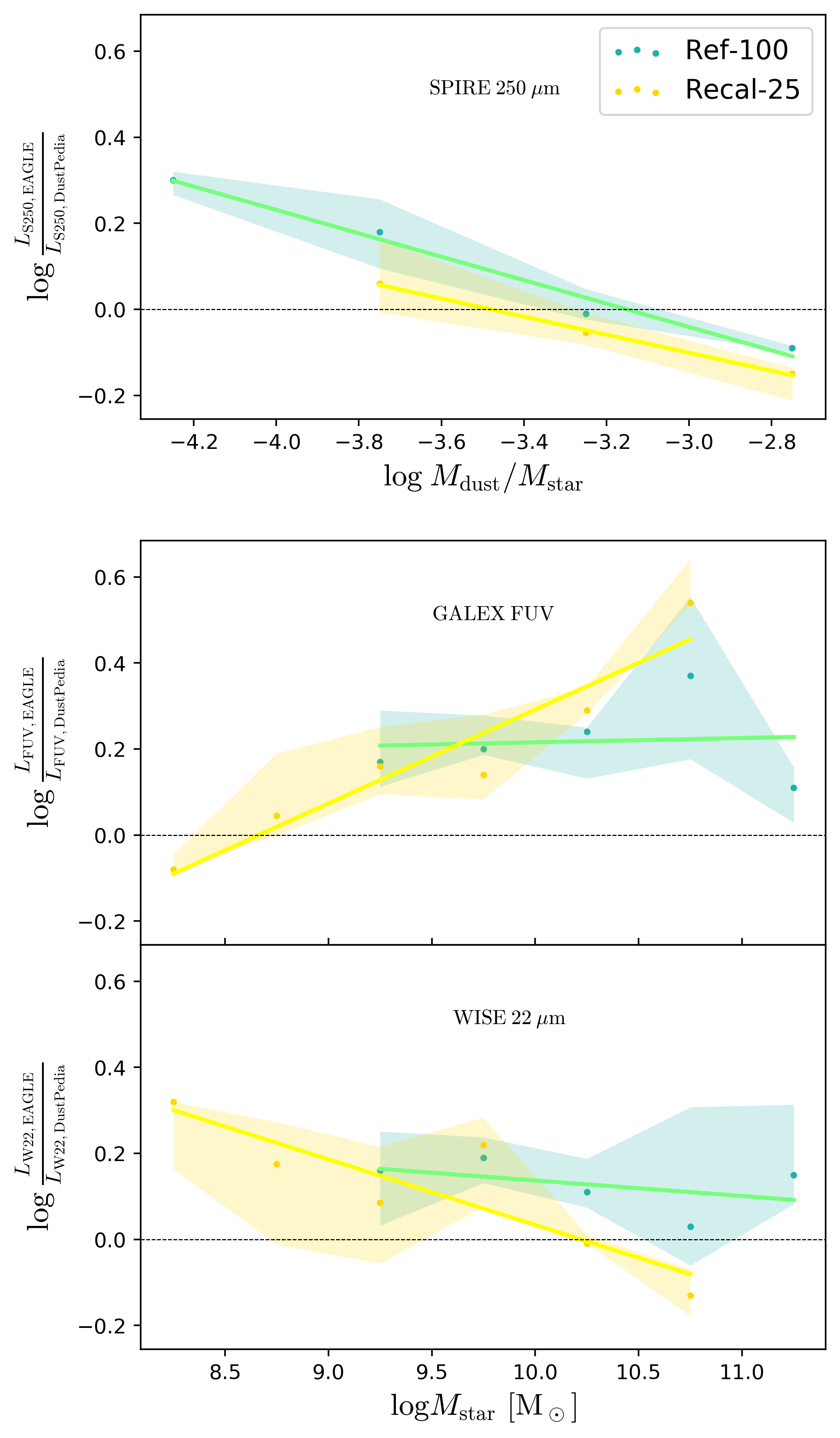}
    \caption{Top: Differences in the SPIRE \um{250} band between EAGLE and DustPedia as a function of the specific dust mass (x-axis) and the stellar mass (weight in each specific dust mass bin). Middle: Same as top but for the FUV band and as a function of the the stellar mass and sSFR. Bottom: Similar as above but for the WISE \um{22} band. The lines represent the linear fit, the shaded regions show $16-84\%$ range.}
    \label{fig:del_line}
\end{figure}

We compute the ratio of luminosity in a band for the EAGLE samples to  luminosity in the same band for the DustPedia sample, in narrow 2D bins of $M_{\mathrm{star}}$ and sSFR or of specific dust mass and $M_{\mathrm{star}}$. The binning minimises the effect of the different galaxy mixture in the different samples. We use the luminosity values from the \textsc{cigale} fits for all three samples and again we exclude $10\%$ of the galaxies with the most irregular SEDs. We have investigated differences in different bands, from FUV to \um{250}, and their potential correlation with stellar mass, (specific) dust mass and (specific) SFR, however we only show those that reveal clear trends. 
The results are presented in Fig. \ref{fig:del_line}. Each point in the specific dust mass bin (top panel) is a median of those in different $M_{\mathrm{star}}$ bins. Each point in the stellar mass bin (middle and bottom panels) is a median of those in different sSFR bins.

From the top panel, it is evident that deviations in the SPIRE \um{250} band correlate with the specific dust mass. The decreasing slope indicates that for the low specific dust mass galaxies our models over-predict radiation in the SPIRE \um{250} band and contrarily, the model slightly under-predicts the \um{250} luminosity for the very dusty galaxies. This discrepancy may be a symptom of the constant dust-to-metal ratio we assume for all EAGLE galaxies (see Sect. \ref{sec:SKIRT}), since the diffuse dust dominates at these wavelengths. 
In particular, it is evident that simply increasing (decreasing) the global dust-to-metal ratio will not eliminate the difference in the slope of the lines in the figure. Recent studies report correlations of the dust-to-metal ratio with galaxy properties like stellar mass and metallicity \citep{DeVis2019,Li2019, Lagos2019}, however it remains to be seen if these correlations are strong enough in the stellar mass regime we consider here. 
Nevertheless, as the dust from the birth clouds can also contribute at these wavelengths, we argue that a modification of the implementation of the subgrid star-forming regions is required as well.

We detected opposite trends with $M_{\mathrm{star}}$ for luminosities in the FUV and the WISE \um{22} bands, as shown in Fig. \ref{fig:del_line}, bottom two panels. 
The correlation is stronger for Recal-25 than for the lower resolution Ref-100. 
These trends reveal that the FUV and MIR emissions are coupled, which is expected since more attenuation of the stellar light in the FUV implies more dust emission in the MIR. Interestingly, neither Fig. \ref{fig:seds_sub} nor \ref{fig:SEDs_bins} unveils this connection - the figures only indicate an excess in both parts of the spectrum. Ideally, the changes in representation of the star-forming regions should be such that, in the lowest mass galaxies, the FUV emission remains unchanged whereas the MIR emission  decreases by $\sim 0.3 ~\mathrm{dex}$. 
For the most massive galaxies, the FUV attenuation should increase substantially, without a major change in the MIR emission. Because the FUV emission predominantly originates from young and massive stars and WISE \um{22} from dust heated by these stars, these bands are severely affected by the geometry of the star-forming regions. Improvement would be expected if the most massive stars in the star-forming regions would have a higher $f_{\mathrm{PDR}}$, and vice versa for the less massive. Whether this adjustment of the geometry is possible with \textsc{mappings-iii} templates, it is not clear at this stage.

Our analysis demonstrates that further study on the modelling of the star-forming regions is needed. There are several approaches that can be taken to tackle this problem. The use of the \textsc{mappings-iii} templates can be revisited, however with improvement of the original procedure by applying additional constraints based on the results from this study. 
For instance, the dust-to-metal fraction, currently a constant parameter in the procedure, can be modified to a variable one. We will explore these avenues in future work. The forthcoming cosmological simulations can benefit from these types of post-processing procedures; they can improve the calibration of the subgrid parameters of the simulations, since the comparison with the observations could then be implemented directly in the observational, i.e. flux, space.

\section{Summary}
\label{sec:sum}

In this study, we investigated across a broad wavelength range whether the EAGLE simulations combined with \textsc{skirt}, provide a proper description of galaxies in the Local Universe. We compare them with the DustPedia sample in a consistent manner using the SED fitting tool \textsc{cigale} to derive all physical properties for all the samples. Though galaxies have very different distributions in individual quantities, the relations between these quantities broadly hold, but a number of discrepancies are present as well:

- Comparing scaling relations between luminosities directly from the databases, reveals discrepancies in the MIR, with much better agreement in the optical and FIR range. Similar results are obtained comparing fitted SEDs and the physical property-luminosity proxy relations, with additional finding of deviations in the UV range.

- Scaling relations between the physical properties show that those relations that are dependent on the global energy, diffuse dust and stellar mass show satisfactory agreement, while the relations and SED regimes primarily driven by the properties of the star-forming regions show discordance.

- To understand the origin of these discrepancies, we analyse only the star-forming galaxies, applying the $\mathrm{sSFR}>-10.8 ~\mathrm{yr^{-1}}$ threshold. Most of the relations improved significantly, indicating the importance of the difference in the galaxy population mix between the samples.

- An analysis of the $\mathrm{IRX}-\beta$ relation, despite the great overall agreement, shows discrepancies which are mainly caused by the limitations in the subgrid treatment of the star-forming regions.

- We quantify the deviations in the median SEDs, and their correlation with galaxy properties. We find trends that can help to improve and optimise the future re-calibration process, necessary for the more realistic modelling of the H\textsc{ii} regions. 

- This detailed comparison highlights the successes and shortcomings of the current panchromatic modelling. This new knowledge indicates the areas where the procedure can be improved, with the aim to implement it in the future cosmological simulations to assist in their calibration process.

\section*{Acknowledgements}
DustPedia is a collaborative focused research project supported by the European Union under the Seventh Framework Programme (2007-2013) call (proposal no. 606847). We acknowledge the Virgo Consortium for making EAGLE simulation data available. 

This research made extensive use of the NumPy, MatPlotLib and Pandas Python packages.




\bibliographystyle{mnras}
\bibliography{bib} 




\appendix
\section{The mock checks for the EAGLE galaxies}
\label{appendix:app1}
\begin{figure}
	\includegraphics[width=\columnwidth]{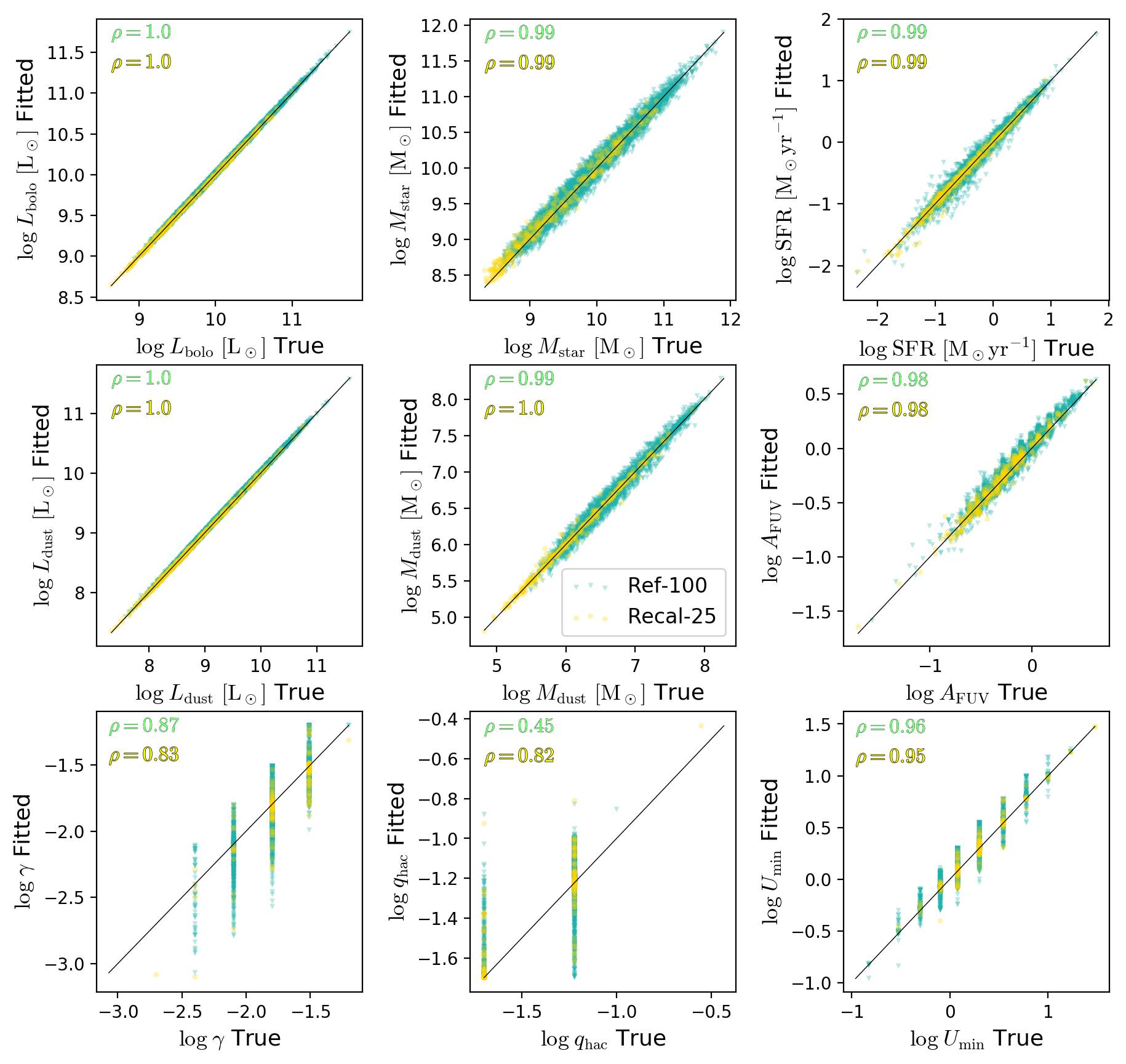}
    \caption{Results of the mock analysis by \textsc{cigale} for the two EAGLE samples. The Spearman coefficients are shown in the corners. The black line is one-to-one relation.}
    \label{fig:app1}
\end{figure}

In this section, we perform the mock analysis which implies running additional \textsc{cigale} module that derives mock fluxes for each galaxy based on their respective best fit. Each flux is then varied in order to introduce noise, and these mock observations are then fitted again to derive the physical properties. In Fig. \ref{fig:app1} results for nine properties are presented. The bottom row represents parameters associated with the THEMIS dust model: $\gamma$ is the fraction of the dust luminosity originating from photo-dissociation regions,  $q_{hac}$ is the fraction of the total dust mass that is in small hydrocarbon grains, and $U_{\rm min}$ is the minimum intensity of the stellar radiation necessary the heat the dust grains. 

The highest deviation is seen for $q_{hac}$, as found for the DustPedia sample \citep{Nersesian2019}. The rest of the parameters agree remarkably well, with the higher Spearman coefficients compared to the DustPedia sample (see Fig. B.1 in \citet{Nersesian2019}). The origin of the better correlation is the completeness of the flux datasets in the EAGLE-\textsc{skirt} database (29 bands for all galaxies), contrarily to DustPedia where the median number of bands per galaxy is 20 (see Table \ref{tab:DE}).

\bsp	
\label{lastpage}
\end{document}